\documentclass{aastex63}
\usepackage{graphicx}
\usepackage{CJK}

\received{----}
\revised{---}
\accepted{November 22, 2020}
\submitjournal{ApJ}

\begin{document}
\begin{CJK*}{UTF8}{gbsn}

\title{Red Supergiants in M31 and M33 \\
I. The Complete Sample}

\correspondingauthor{Biwei Jiang}
\email{bjiang@bnu.edu.cn}

\author[0000-0003-1218-8699]{Yi Ren (任逸)}
\affiliation{Department of Astronomy, Beijing Normal University, Beijing 100875, People's Republic of China}

\author[0000-0003-3168-2617]{Biwei Jiang (姜碧沩)}
\affiliation{Department of Astronomy, Beijing Normal University, Beijing 100875, People's Republic of China}

\author[0000-0001-8247-4936]{Ming Yang (杨明)}
\affiliation{IAASARS, National Observatory of Athens, Vas. Pavlou and I. Metaxa, Penteli 15236, Greece}

\author[0000-0001-5197-4858]{Tianding Wang (王天丁)}
\affiliation{Department of Astronomy, Beijing Normal University, Beijing 100875, People's Republic of China}

\author[0000-0002-5649-7461]{Mingjie Jian (简明杰)}
\affiliation{Department of Astronomy, School of Science, The University of Tokyo, 7-3-1 Hongo, Bunkyo-ku, Tokyo 113-0033, Japan}

\author[0000-0003-4988-3513]{Tongtian Ren (任桐田)}
\affiliation{Department of Astronomy, Beijing Normal University, Beijing 100875, People's Republic of China}



\begin{abstract}

The aim of this paper is to establish a complete sample of red supergiants (RSGs) in M31 and M33. The member stars of the two galaxies are selected from the near-infrared (NIR) point sources after removing the foreground dwarfs from their obvious branch in the $J-H/H-K$ diagram with the archival photometric data taken by the UKIRT/WFCAM. This separation by NIR colors of dwarfs from giants is confirmed by the optical/infrared color-color diagrams ($r-z/z-H$ and $B-V/V-R$), and the Gaia measurement of parallax and proper motion. The RSGs are then identified by their outstanding location in the members' $J-K/K$ diagram due to high luminosity and low effective temperature. The resultant sample has 5,498 and 3,055 RSGs in M31 and M33 respectively, which should be complete because the lower limiting $K$ magnitude of RSGs in both cases is brighter than the complete magnitude of the UKIRT photometry. Analysis of the control fields finds that the pollution rate in the RSGs sample is less than 1\%. The by-product is the complete sample of oxygen-rich asymptotic giant branch stars (AGBs), carbon-rich AGBs, thermally pulsing AGBs and extreme AGBs. In addition, the tip-RGB is determined together with its implication on the distance modulus to M31 and M33.

\end{abstract}



\section{Introduction} \label{sec:intro}
The red supergiants (RSGs) are Population I massive stars in the core-helium burning stage. It is generally believed that the initial mass of a RSG is at least $\sim 8 M_{\odot}$. But the lower limit of initial mass for the RSG population may be as low as $7M_{\odot}$ or even $6M_{\odot}$ \citep{2019AandA...629A..91Y}. The radius of the RSGs can reach $\sim 1500 R_{\odot}$ \citep{2005ApJ...628..973L}, and they have low surface gravity and high luminosity of $~3,500-630,000 L_{\odot}$ \citep{2008IAUS..250...97M,2016ApJ...826..224M}. A complete catalog of RSGs is the basis to study the properties of RSGs more accurately, such as to examine massive star evolution as a function of metallicity \citep{1980A&A....90L..17M,2002ApJS..141...81M,2013NewAR..57...14M}, to estimate the total contribution of dust by RSGs to interstellar dust \citep{1975MSRSL...8..369R,1978A&A....70..227K,2016ApJ...825...50G}, and to calibrate the period-luminosity (P-L) relations of RSGs \citep{2006MNRAS.372.1721K,2011ApJ...727...53Y,2012ApJ...754...35Y,2018ApJ...859...73S,2019MNRAS.487.4832C,2019ApJS..241...35R} and the scaling relations between granulation and stellar parameters \citep{2020ApJ...898...24R}.

The Small Magellanic Cloud (SMC), Large Magellanic Cloud (LMC), Triangulum Galaxy (M33) and Andromeda Galaxy (M31) are all nearby that RSGs can be detected and resolved individually, which provide important cases to learn the statistical properties of RSGs in a galaxy. There have been some collections of RSG samples in these galaxies.  The sample of RSGs was on the scale of a few tens objects in the early studies of LMC and SMC \citep{1980MNRAS.193..377F,1981MNRAS.197..385C,1983ApJ...272...99W,2000MNRAS.313..271P}, and increased to a couple of hundreds later \citep{2002ApJS..141...81M,2003AJ....126.2867M,2012ApJ...749..177N,2015AandA...578A...3G,2011ApJ...727...53Y,2012ApJ...754...35Y}. Recently, \citet{2019AandA...629A..91Y,2020AandA...639A.116Y} identified 1,405 and 2,974 RSGs in SMC and LMC respectively, which is a drastic increase from previous studies and estimated to be about 90\% complete. This revolutionary progress comes from both much more collections of data and the method to remove the foreground stars. Specifically, they combine a variety of color-magnitude diagrams (CMDs) to identify RSGs and remove the foreground contamination by Gaia's proper motion and parallax.

M31 and M33 are much more distant than the MCs with distance moduli larger by about 5 magnitudes, which calls for alternative methods to identify RSGs. Previously, \citet{2006AJ....131.2478M,2007AJ....133.2393M,2009ApJ...703..420M,2012ApJ...750...97D} selected the initial RSGs sample by $V<20$ and $V-R\ge0.85$ for M31 and $V-R\ge0.6$ for M33 respectively by the Local Group Galaxies Survey (LGGS) observation. The $V$ band criteria was set to ensure sufficient brightness to avoid confusion with asymptotic giant branch stars (AGBs) and the color index was set to limit the stars to a K- and later type. The foreground dwarfs were further removed from this initial sample by the $B-V/V-R$ diagram \citep{2009ApJ...703..420M,2012ApJ...750...97D}. Consequently, they identified 437 and 776 RSG candidates in M31 and M33 respectively. \citet{2016ApJ...826..224M} measured radial velocities and determined spectral types for 255 (about 60\%) of these stars and confirmed they are truly RSGs after comparing their radial velocities  with the expected values of the Population I objects in M31 \citep{2009ApJ...703..420M}.

The RSGs samples in M31 and M33 identified by \citet{2009ApJ...703..420M} are far from complete.  \citet{2019ApJS..241...35R} found that the minimum luminosity of RSGs in these samples is about 1 mag above the theoretical limit of RSGs for a 9 $M_{\odot}$ star, which indicates that the samples missed the faint RSGs. In addition, \citet{2019AJ....158...20M} discovered 889 RSG candidates in the Milky Way galaxy from Gaia Data Release 2 (DR2). Considering that both M31 and M33 are spiral galaxies like the Milky Way, the number of RSGs should be comparable. Moreover, the sample of 889 RSGs in the Milky Way cannot be complete. In comparison with the newly found large sample (a couple of thousands) of RSGs in SMC and LMC, M31 and M33 with a much larger geometrical size are expected to host many more RSGs.  As mentioned in \citet{2016ApJ...826..224M}, their completeness limit was set by $V\leq20$,  which implies that the RSGs population of M31 was only complete down to $\sim15M_{\odot}$, corresponding to $\log L/L_{\odot}\sim4.7$ according to the mass-luminosity relation of massive stars ($L/L_{\odot}=(M/M_{\odot})^{4}$, \citealt{1971A&A....10..290S}). If we set out to complete the identification of RSGs in M31 and M33 down to $\sim7M_{\odot}$, i.e. the bolometric magnitudes $M_\mathrm{bol}$ of $\sim-3.71$ mag, the $V$ band magnitudes will be 22.63 and 22.89 for M31 and M33 respectively by taking the distance modulus of 24.40 \citep{2009A&A...507.1375P} and 24.66 \citep{2007Natur.449..872O}, the bolometric correction $\mathrm{BC}_{V}$ of -0.94 for a 4000K RSG \citep{2009ApJ...703..420M}, and the extinction $A_{V}=1$ \citep{2009ApJ...703..420M}.  With the increase of the photometric error to the faint objects, it becomes more and more difficult to distinguish the foreground dwarfs from the member RSGs by using the $B-V/V-R$ diagram. Thus it is hard to yield a complete and pure identification of faint, cool RSGs in M31 and M33 by using the optical data like LGGS.

In this work we try to establish a complete sample of RSGs in M31 and M33 using near-infrared data in a new way. The paper is organized as Section \ref{sec:data} on the data, Section \ref{sec:Removing the foreground stars} on the method to remove the foreground stars, and Section \ref{sec:Identifying the RSGs in the CMD} on how to identify RSGs.

\section{Data and Reduction} \label{sec:data}
The $JHK$ brightness comes from images taken with the Wide Field Camera (WFCAM) from mid 2005 to 2008 on the 3.8 m United Kingdom Infra-Red Telescope (UKIRT) located in Hawaii \citep{2013ASSP...37..229I}. WFCAM consists of four Rockwell Hawaii-II (HgCdTe $2048 \times 2048$) detectors, each covering $13'.65$ on sky with $0.4''/$pixel. For some exposures with microstepping, which is used to recover some of the lost resolution when observing conditions are undersampled, the microstepped frames are interwoven to give an effective sampling of $0.2''/$pixel in the $2\times2$ microstep mode or $0.133''/$pixel in the $3\times3$ microstep mode\footnote{http://casu.ast.cam.ac.uk/surveys-projects/wfcam/technical/interleaving}. For the images we used in this work, the average seeing on all frames varied between $\sim$ $0.7''-1.2''$. The images were processed by the Cambridge Astronomical Survey Unit (CASU) and made available via the WFCAM Science Archive\footnote{http://surveys.roe.ac.uk/wsa}. The data products include the calibrated stacked images and the corresponding source catalogs. We further convert the FITS-format source tables into ASCII-format catalogs applying all necessary corrections using program provided by CASU\footnote{http://casu.ast.cam.ac.uk/surveys-projects/software-release/fitsio\_cat\_list.f/view}. The resultant ASCII-format catalogs contain RA, Dec, magnitude, magnitude error, stellar classification flag, etc. We cross-match the results between $J,H$ and $K$ bands with a radius of $1''$.   The sources with the stellar classification flag of -1 (stellar), -2 (probably-stellar), and -7 (source with bad pixels) are kept and regarded to be point source.  For the cases where the flags disagree in the three bands, at least two of the $JHK$ bands must meet the above conditions to be selected. We add a ``N\_Flag" index to the $JHK$ catalogs to indicate the number of bands in which the source is identified as a stellar source, i.e. 3 means all three bands are labelled "stellar". Finally there are 1,245,930 and 203,486 point sources in M31 and M33 respectively. Different from other works, the source with a flag of -7 is kept in order to guarantee the completeness of the sample. For M31, the percentage of ``-7'' sources is about 5\% in the $J$ band, while up to about 40\% in the $H$ and $K$ band in the initial sample. The large fraction of ``-7'' sources in the $HK$ band and difference with $J$ band  is caused by that the $H$ and $K$ bands data used were all taken  in the $3\times3$ microstep mode, while all of the $J$ band data is taken in the non-microstep mode. As a result, the percentage of ``-7'' sources increases several times in the $H$ and $K$ bands because the data is flagged as having bad pixels in the core radius. The case of M33 is similar. We checked the original source catalogs and found that most of the sources with a fractional number of bad pixels is due to soft-edged aperture. So the photometry of ``-7'' sources are also considered reliable.

In order to examine the photometry accuracy of UKIRT, the $J$ and $K$ magnitudes from UKIRT are compared with those of the Two Mircon All-Sky Survey (2MASS; \citealp{2006AJ....131.1163S}). Since  the UKIRT astrometry and photometry are calibrated with the 2MASS point source catalog, the UKIRT photometric results agree very well with the 2MASS as expected.
In the case of M31, the 2MASS 6x point source catalog can reach to $J \sim 20$ mag and $K \sim 18.5$ mag, while the sources flagged ``AAA"\footnote{``AAA'' represents sources with signal-to-noise ratio $>10$ and magnitude uncertainty $<0.10857$ in $JHK$ bands} reach to $J \sim 17.5$ mag and $K \sim 16$ mag. The $1 \sigma$ differences between 2MASS and UKIRT are $\sim 0.2$ mag at $J \sim17.5$ mag and $K \sim 16.5$ mag. In the case of M33, the main 2MASS point source catalog can reach to $J \sim17.5$ mag and $K \sim 16.5$ mag, while the ``AAA" sources to $J \sim 16$ mag and $K \sim 15$ mag with the $1 \sigma$ difference better than 0.1 mag. These differences are well within the claimed uncertainty.

\begin{figure}
	\centering
    \includegraphics[scale=0.6]{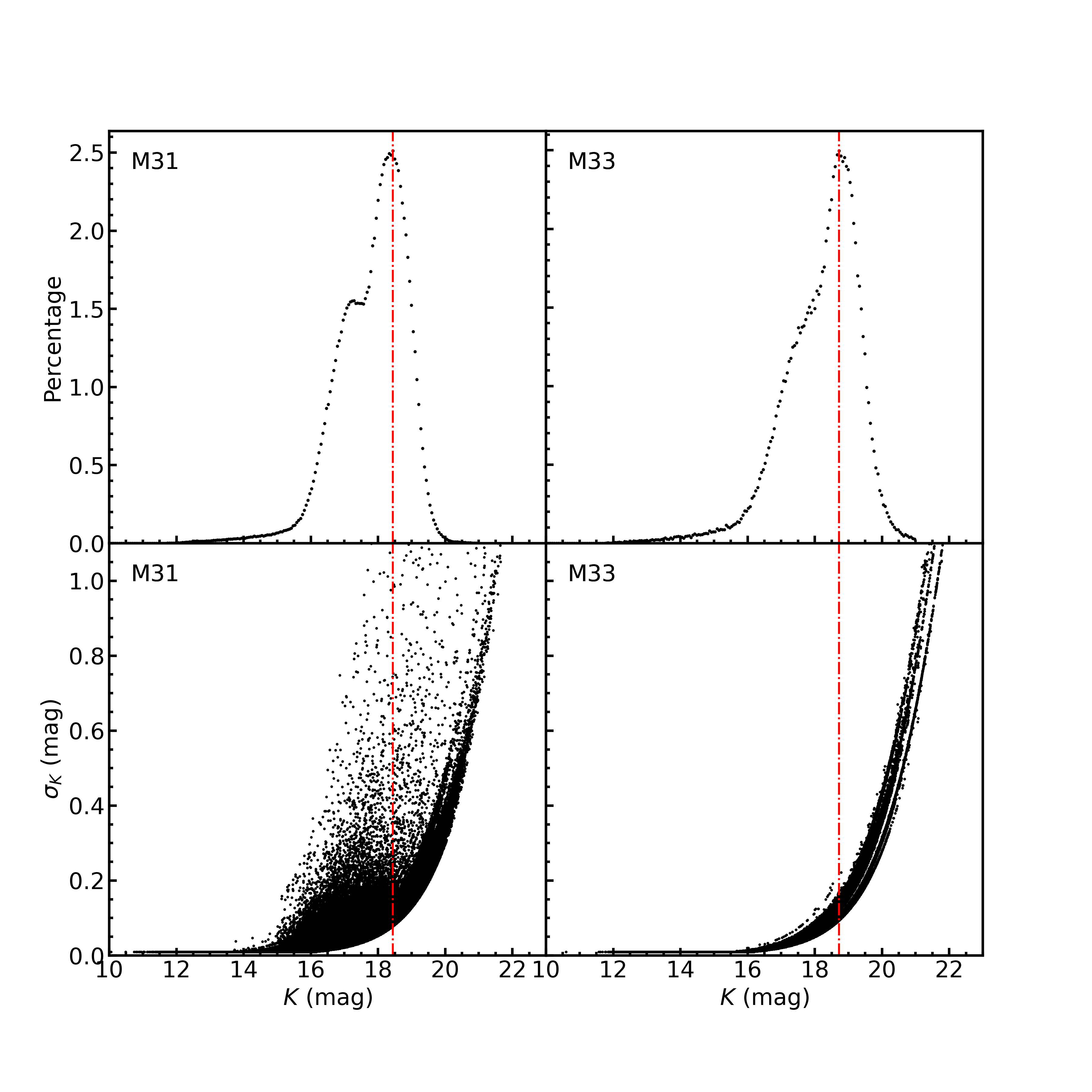}
	\caption{The distribution of the $K$ magnitude (upper panels) and its error (lower panels) in M31 (left) and M33 (right) regions. A drop-off of 18.44 mag and 18.72 mag is indicated by the red dot-dash lines.  \label{fig:K_Kerr}}
\end{figure}

\begin{figure}
	\centering
    \includegraphics[scale=0.4]{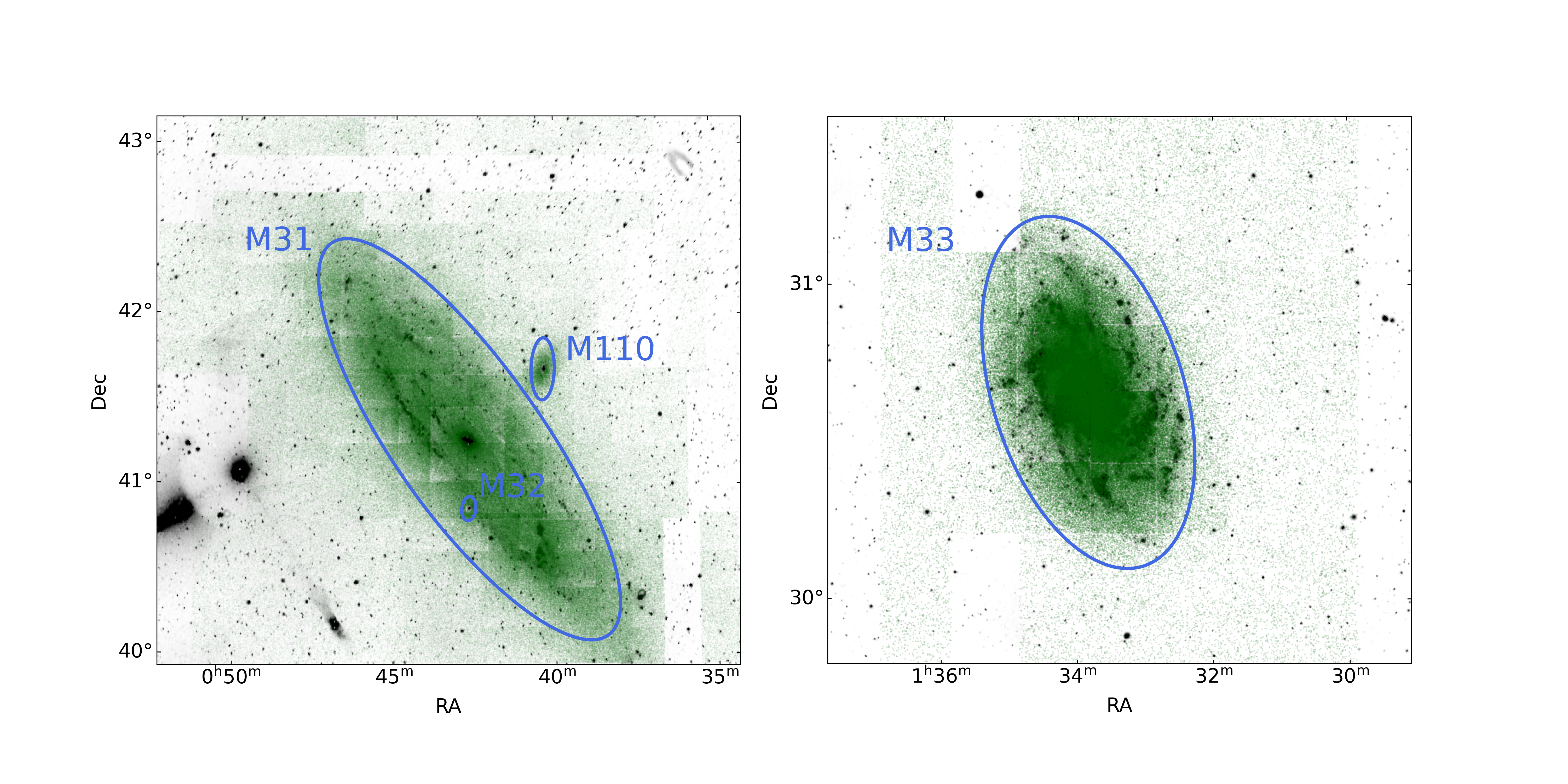}
	\caption{The areas observed by the UKIRT/WFCAM in M31 (left) and M33 (right), where the green dots decode the point sources and the background image is taken from the GALEX ultraviolet survey. The adopted boundary of every galaxy including M32 and M110 is encircled by ellipse.  \label{fig:distribution_field}}
\end{figure}

\begin{figure}
	\centering
    \includegraphics[scale=0.45]{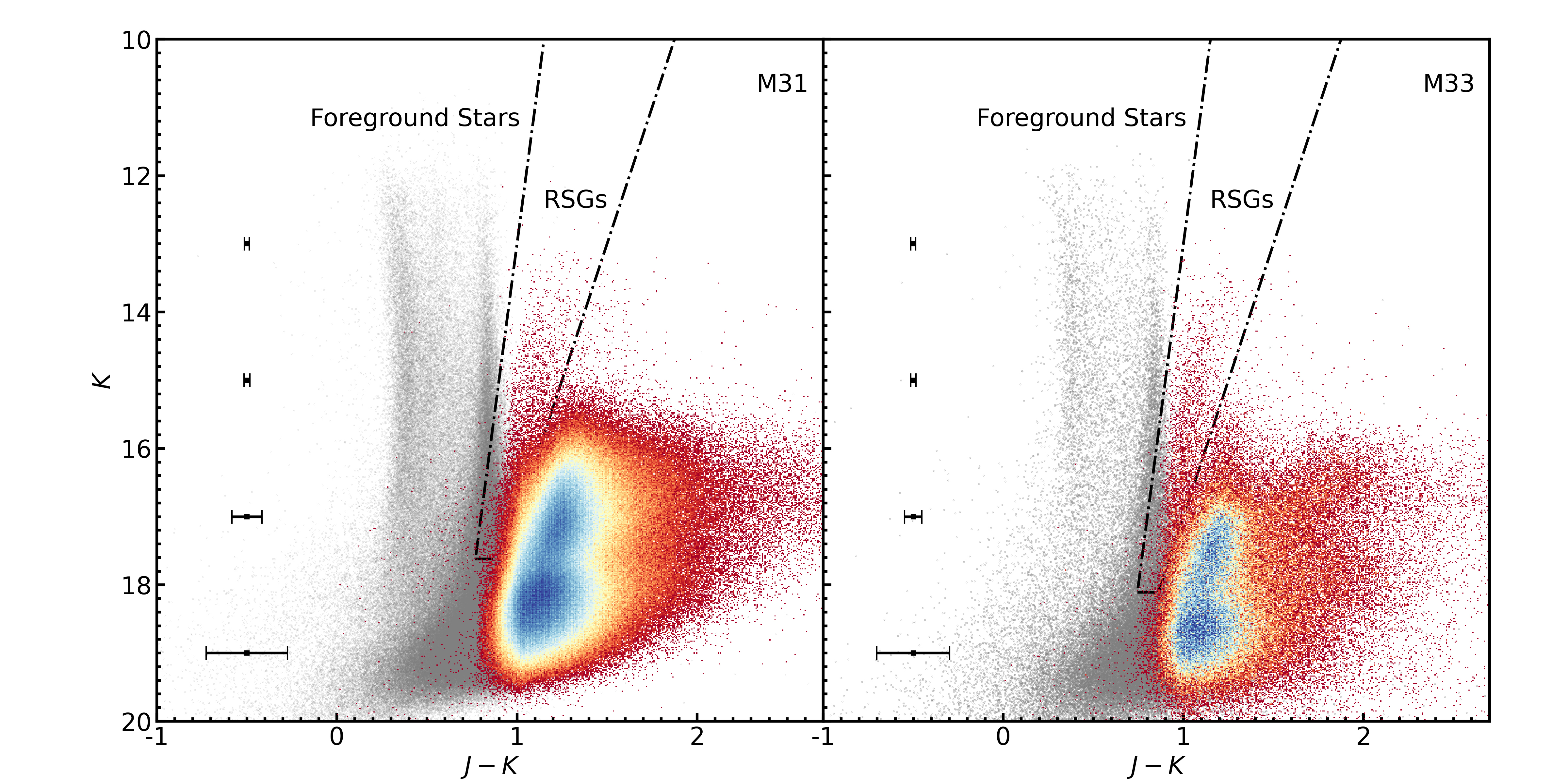}
	\caption{The observed color-magnitude diagram of the initial sample in the M31 (left) and M33 (right) fields. The member stars are colored. The error bars show the mean error of $J-K$ at different $K$ magnitudes. \label{fig:fgd_member.png}}
\end{figure}

The distributions of $K$ band magnitude and corresponding error are shown in Figure \ref{fig:K_Kerr}. In general, the photometry is better for M33 than M31 in both depth and accuracy because the star field in M33 is less crowded. If the magnitude with 0.5 mag brighter than the drop-off point is considered to be complete, then the sources brighter than 17.94 and 18.22 are complete in M31 and M33 respectively. The spatial locations of the selected stars are shown in Figure \ref{fig:distribution_field}, where M32 and M110 are labelled as well by their location and size. It can be seen that the samples cover almost all the regions of M31 and M33, with a very small part of M33 missed. There are some additional fields beyond M31 and M33 which will be used as control fields to estimate the pollution rate of RSGs. The observed CMDs of the initial sample is shown in Figure \ref{fig:fgd_member.png}, where the lately identified member stars are decoded by color.

\section{Removing the foreground stars} \label{sec:Removing the foreground stars}
%%\subsection{Previous Methods} \label{sec:Previous Methods}

Although M31 and M33 are not located in the Galactic plane, \cite{2007AJ....133.2393M} showed that the contamination by foreground stars is serious towards these sightlines. There are a few ways to remove foreground stars. \citet{2019AandA...629A..91Y, 2020AandA...639A.116Y,2020.LMC.Y} separate efficiently the SMC and LMC members from foreground stars by using astrometric solution from Gaia/DR2 since the MC members concentrate on the values expected from the motion of MCs relative to the Galaxy. Unfortunately this method cannot be applied to M31 and M33 effectively because they are so distant that the proper motions and parallaxes are too small to be measurable by Gaia. In the sample, there are only 6.7\% ($\sim 83,260/1,245,930$) and 9.3\% ($\sim 18,901/203,486$) sources with Gaia parallaxes and proper motions measurements and many of which are unreliable. For M31 and M33, \citet{2016ApJ...826..224M} removed foreground stars by radial velocities and spectral type from optical spectroscopy, which can identify RSGs and membership correctly but only for very bright sources. As mentioned earlier, the optical $B-V/V-R$ diagram is used to remove foreground dwarfs.  This two-color method is deeper than spectroscopy and work very well to distinguish foreground dwarfs from RSGs. But this method is useful only for high-accuracy photometry and limited to bright sources (c.f. \citealt{2009ApJ...703..420M,2012ApJ...750...97D,2016ApJ...826..224M} and our later analysis).

Instead of optical observation, we rely mainly on the near-infrared photometric data. Because the effective temperature of RSGs ranges in 3000-5000\,K \citep{2008IAUS..250...97M,2010ApJ...719.1784N,2019AandA...629A..91Y}, their major radiation goes around near-infrared so that RSGs shall be most easily detectable in the $J$, $H$ and $K$ bands. For a RSG with color index $V-K \sim 4.0$, its $K$ magnitude would be about 18 at $V=22$. The infrared band also has much less extinction than the optical \citep{2019ApJ...877..116W}. We take advantages of near-infrared bands to identify  RSGs by using the UKIRT observation of M31 and M33. Besides, the Gaia DR2 data is used to remove the foreground giant stars, though practically no object is removed in this way.

\subsection{Removing the Foreground Dwarfs} \label{sec:Removing the Foreground Dwarfs}
\subsubsection{By the Near-Infrared Color-Color Diagram} \label{sec:By the Near-Infrared Color-Color Diagram}

We remove the foreground dwarfs by the near-infrared color-color diagram. The study of \citet{1988PASP..100.1134B} found that the intrinsic color indexes have clear bifurcations on the $J-H/H-K$ two-color diagram for giant and dwarfs. Dwarfs have higher surface gravity than giants or supergiants, the collision rate between atoms is higher and molecules are easier to form \citep{1995ApJ...445..433A}. This makes molecules form at relatively high temperatures in dwarfs, causing absorption in the $H$ band and darkening the $H$-band brightness and eventually leading to smaller $J-H$ and bigger $H-K$ than giants.

\begin{figure}
	\centering
    \includegraphics[scale=0.45]{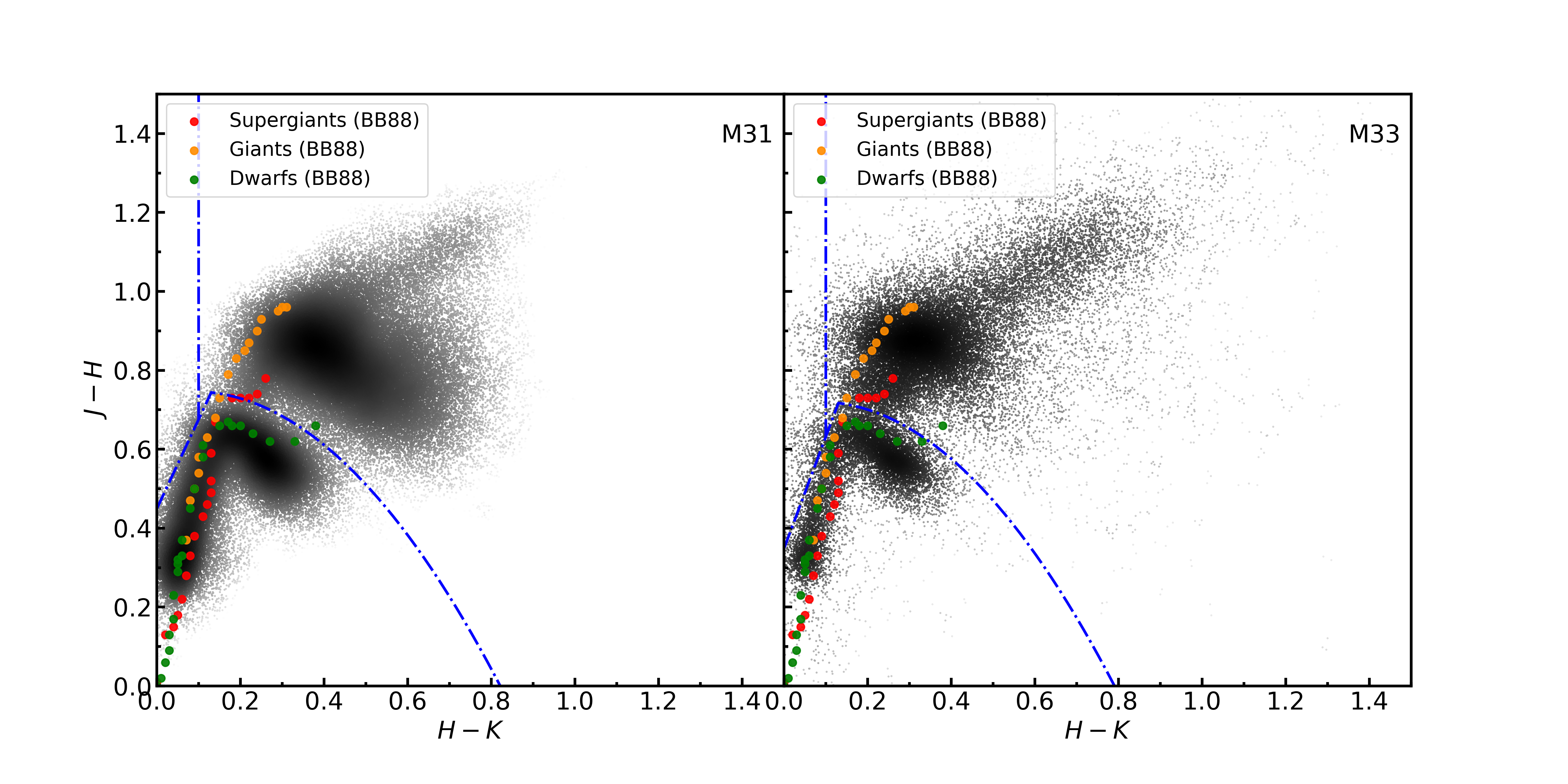}
	\caption{The $J-H/H-K$ diagram of the good-quality measurements with ``N\_Flag=3" and the $JHK$ photometric errors less than 0.05 mag. The colors represent the surface density. The criteria to remove foreground dwarfs are shown by dot-dash lines, which is compared with the intrinsic color indexes of dwarfs (grey dots), giants (green dots) and supergiants (blue dots) by \citet[BB88 for short]{1988PASP..100.1134B}.  \label{fig:UKIRT-Criteria}}
\end{figure}

The borderline between dwarfs and giants are re-defined intentionally and specifically though \citet{1988PASP..100.1134B} already obtained the intrinsic color indexes. For this purpose, the high-accuracy photometric data are chosen  with the error of $J$, $H$ and $K$ band photometry less than 0.05\,mag and N\_Flag = 3 which means the object is identified as "stellar" in all three bands. The $J-H / H-K$ diagrams for these accurate photometries are shown in Figure \ref{fig:UKIRT-Criteria} for M31 and M33. It can be seen that there is a very clear boundary between giants and dwarfs. With the increase of $H-K$, $J-H$ increases to about 0.7 and then turns down for dwarfs, meanwhile $J-H$ of red giants  and supergiants starts from about 0.7 and increases monotonically. The trends and values coincide very well with the result of \citet{1988PASP..100.1134B}.

The dividing line is defined quantitatively. In a step of 0.01 in the range of $H-K$ from 0.05 to 0.3, the point of the maximum surface density is calculated on the dwarf branch, and then the piecewise function is used to fit those points. When $H-K \leq 0.13$, the function is linear; for $H-K > 0.13$, the function is quadratic. This piecewise function represents the relation of $J-H$ and $H-K$ of dwarfs. Since the foreground extinction is pretty small (an average $A_{V}\sim0.17$ \citep{1998ApJ...500..525S,2011ApJ...737..103S,2011ApJ...726...39C}), these colors basically represent the intrinsic color indexes of dwarfs. We shift the piecewise functions up by 0.12 and 0.09 mag with eye-check and take them as the dividing lines between giants and dwarfs in the M31 and M33 fields, where the difference of shift in the vertical axis is caused by different foreground extinction to the two galaxies, i.e. $A_{\rm V} \sim$ 0.17 mag and 0.11 mag for M31 and M33 respectively according to the SFD98 \citep{1998ApJ...500..525S}. The function forms and coefficients of adopted dividing lines are listed in the first rows of Table \ref{tab:dividing lines}.

With the dividing line determined, the criteria is applied to the entire initial sample to remove foreground dwarfs. In addition, we remove sources with $H-K<0.1$ that is apparently bluer than RSGs. This action certainly also removes the blue and yellow supergiants in the galaxy, which are absent in the final catalog of the members in M31 and M33. Finally 414,490 and 77,091 foreground dwarfs are removed in the M31 and M33 fields, i.e. 33.3\% and 37.9\% of the initial samples.

The above method is expected to remove all the foreground dwarfs in the sample, but the uncertainty of color indexes would move some sources around the borderline.  The completeness and pollution rate of the selected member stars are estimated by Monte Carlo simulation. First, the sources with ``N\_Flag=3" and $\sigma_{J,H,K} < 0.05$ mag are taken as a no-error "perfect" sample so that their locations in the $J-H/H-K$ diagram can absolutely decide being a dwarf or a giant. Then, we perform 5000 simulations for a random error with two-dimensional Gaussian distribution for each source whose width is four times the error of $J-H$ and $H-K$.  Finally, the UKIRT/NIR criterion is used to divide dwarfs and giants to compute the completeness and pollution rate of giants when the $JHK$ photometric errors are limited to be less than 0.2 mag. For M31 and M33, the simulation results show that the completeness of the selected stars is about 93\%, and the pollution rate is about 9\%.

Actually, the pollution rate calculated by this method will be overestimated and the completeness will be underestimated. On one hand, we multiply the errors of the sources whose $JHK$ photometric errors are less than 0.05 mag by 4 to simulate the case that the $JHK$ photometric errors are limited to be less than 0.2 mag, which means increasing the photometric errors systemically. On other hand, the distribution of sources with $JHK$ photometric errors less than 0.05 mag is already scattering due to the photometric errors. Therefore, the true pollution rate of the selected giants is smaller than the simulation value, and the completeness is larger than the simulation value.

\subsubsection{Double Check by Optical/Infrared Color-Color Diagrams} \label{sec:Double Check by Optical/Infrared Color-Color Diagrams}

Although the $J-H/H-K$ diagram works very efficiently and is applicable to all the sample stars, the identification deserves to be checked by other methods. One purpose is to confirm the identification, and the other is to further remove some foreground stars which are very close to the borderline in the NIR color-color diagrams (CCD) but may be significantly distinguishable in optical bands in particular for some relatively blue stars.

\begin{figure}
	\centering
    \includegraphics[scale=0.45]{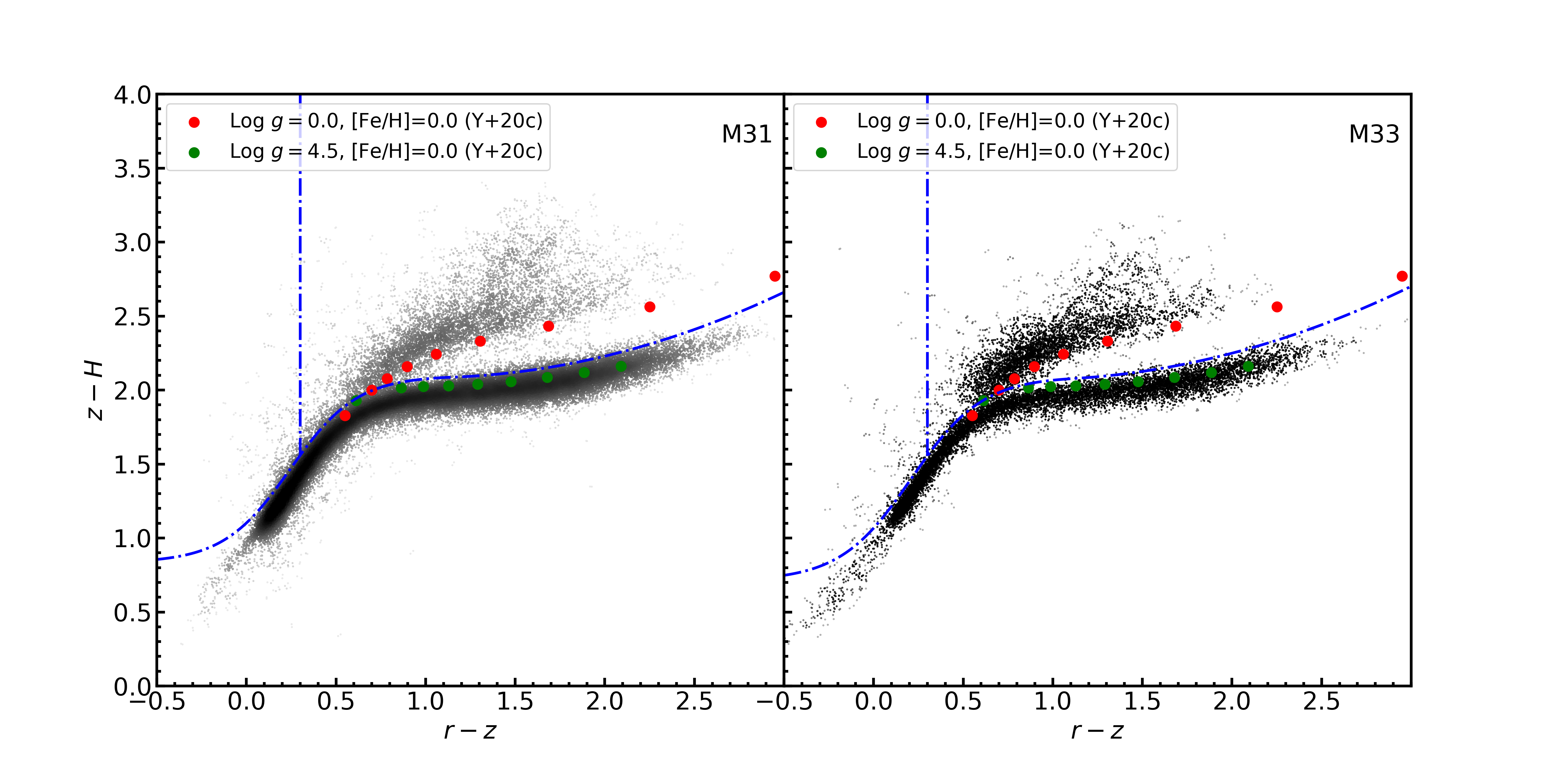}
	\caption{The $r-z/z-H$ diagram of the good-quality measurements with ``N\_Flag=3" and the $rzH$ photometric errors less than 0.05 mag. The symbol convention follows Figure \ref{fig:UKIRT-Criteria}.  The intrinsic color indexes of dwarfs, giants and supergiants from \citet[Y+20c for short]{2020.NGC6822.Y} are shown for comparison.  \label{fig:PS1-Criteria}}
\end{figure}

\begin{figure}
	\centering
    \includegraphics[scale=0.45]{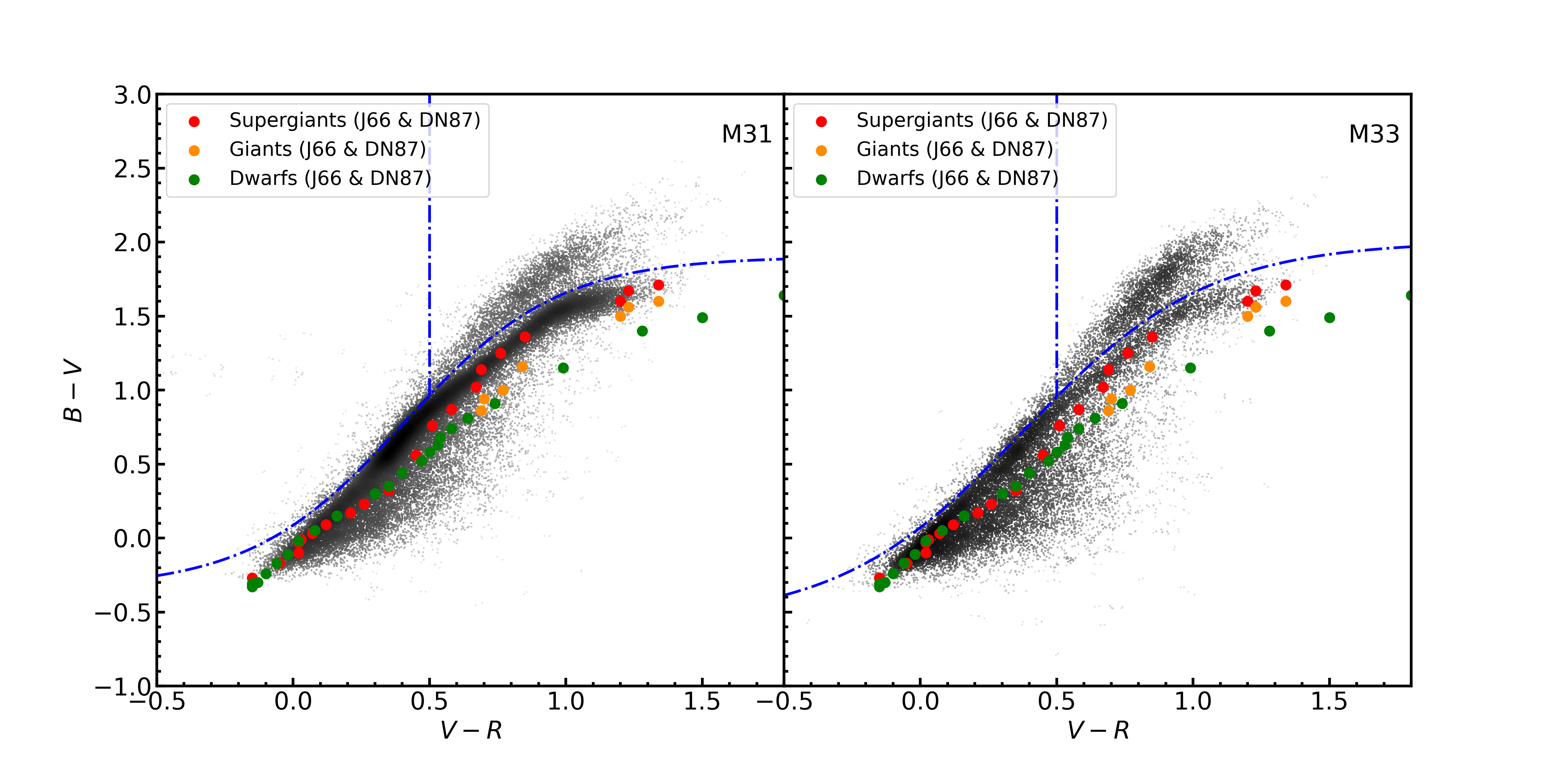}
	\caption{The $B-V/V-R$ diagram of the good-quality measurements with the error of $B-V$ and $V-R$ less than 0.05 mag. The symbol convention follows Figure \ref{fig:UKIRT-Criteria}. The intrinsic color indexes of dwarfs, giants and supergiants obtained by \citet[J66 for short]{1966ARA&A...4..193J}; \citet[DN87 for short]{1987A&A...177..217D} are shown for comparison.  \label{fig:LGGS-Criteria}}
\end{figure}

\begin{figure}
	\centering
    \includegraphics[scale=0.4]{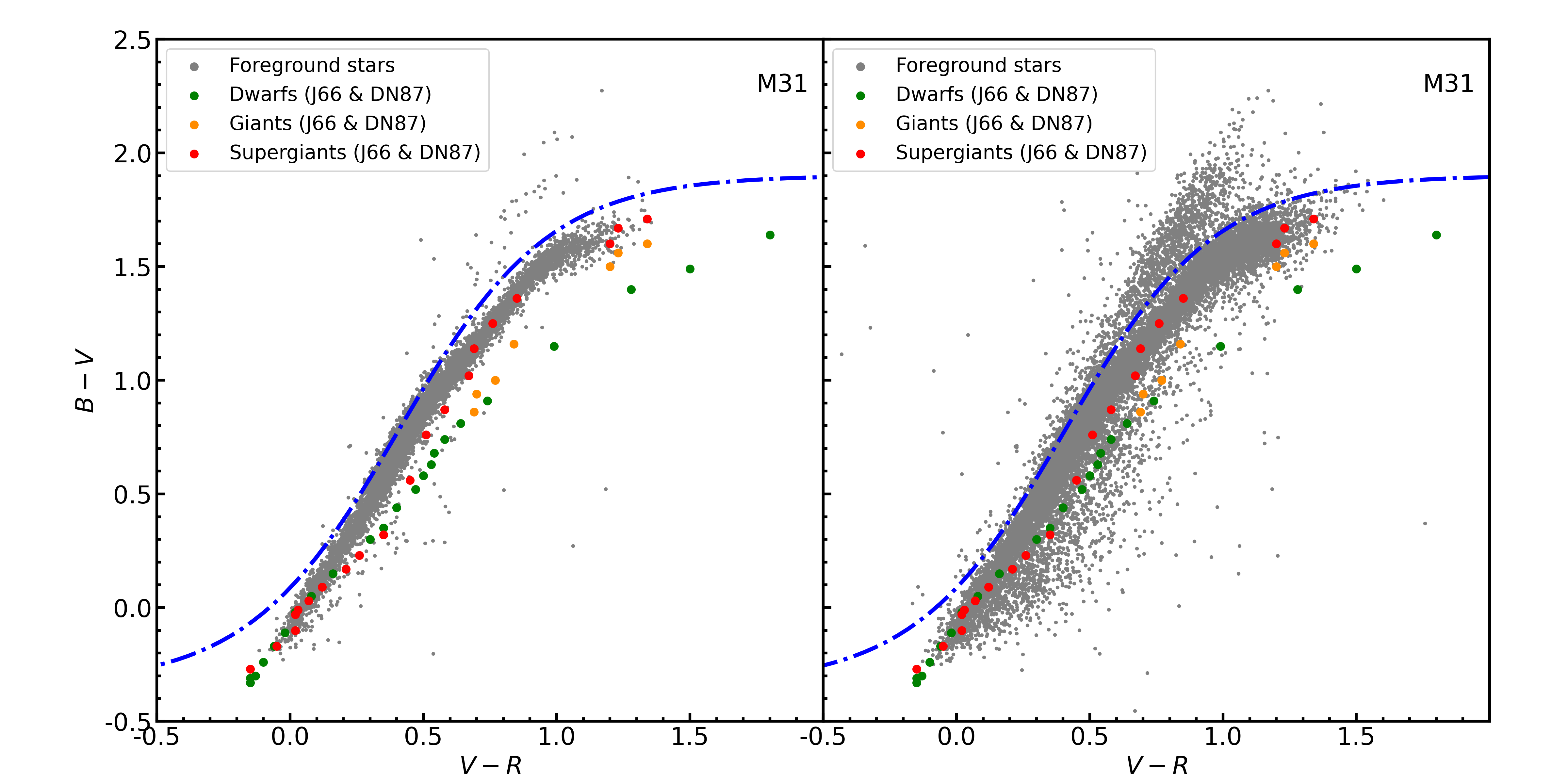}
    \caption{Distribution of all the removed foreground stars in M31 on the $B-V/V-R$ diagram. The left panel shows the foreground stars with the error of $B-V$ and $V-R$ less than 0.01 mag while the right panel for stars with the error of $B-V$ and $V-R$ less than 0.05 mag. For comparison, the intrinsic color indexes of dwarfs, giants and supergiants from \citet{1966ARA&A...4..193J,1987A&A...177..217D} are shown. \label{fig:check_bvr_foreground}}
\end{figure}

\begin{itemize}

\item \textbf{The $r-z/z-H$ diagram.}

By convoluting the spectrum of the MARCS model with transmission functions of $r,z,H$ filters, \citet{2020.NGC6822.Y} found that the $r-z/z-H$ diagram can well distinguish dwarfs from giants. We introduce the $r-z/z-H$ diagram only as an auxiliary method to remove foreground dwarfs, because the information used to distinguish dwarfs from giants here, $H$-band photometry, have already been used in the $J-H/H-K$ CCD (see Section \ref{sec:By the Near-Infrared Color-Color Diagram}), and only those foreground dwarfs with good photometric quality will be removed.
%it is still the $H$ band that plays a role in distinguishing dwarfs from giants

The PS1/DR2 data is used, where the forced mean PSF magnitude is taken for its consideration of both photometric depth and accuracy. The data flags of PS1/DR2 are complex and only those with good measurements are used. The cross-match between UKIRT and PS1 by a radius of $1''$ resulted in 184,750 and 37,304 stars in M31 and M33 with good-photometry. Similar to the method in Section \ref{sec:By the Near-Infrared Color-Color Diagram}, we select the ``good measurement'' sources with $r$, $z$, $H$ band photometric errors less than 0.05\,mag to determine the borderline between dwarfs and giants in the $r-z/z-H$ diagram. The positions of maximum surface density of the dwarf branch on $r-z/z-H$ diagram are calculated in a step of 0.01 of $r-z$ from 0 to 2.5, and then a piecewise function is used to fit those points. When $r-z\leq1.1$, the sigmoid function, which is also known as the Logistic function, is used for fitting, otherwise, the function is a quadratic curve. As shown in Figure \ref{fig:PS1-Criteria},  the piecewise functions are shifted up by 0.12 and 0.10 mag for M31 and M33 to become the dividing lines between giants and dwarfs. It should be noted that a slight difference in the intrinsic color indexes $r-z$ or $z-H$  appears between our results and the MARC model though they are in general agreement.  The functions used  are listed in the second rows of Table \ref{tab:dividing lines}

The criteria is applied to the PS1 sources with ``good measurement" and the photometric error in $r$, $z$, $H$ bands is less than 0.1 mag. The sources below the dividing line or with $r-z < 0.3$ are removed. This removes 102,174 and 17,648 dwarfs in M31 and M33 fields respectively, in which 93,938 (91.9\%) and 15,568 (88.2\%) are also removed by the NIR CCD. In other words, additional 8,236 and 2,080 stars are removed, which are mostly around the borderline in the NIR CCD or in the area close to the center of the galaxy where the photometry is of poor quality.

%PS1 DR2 gives several flags to evaluate the quality of sources, that we take as criteria for rejection. Table \ref{tab:ps1 flags} lists the flag values, with which the sources should be eliminated. And the remainder are considered as ``good-measured" sources.

%The UKIRT catalogs are crossmatched with Gaia data release 2 (DR2), the Panoramic Survey Telescope And Rapid Response System data release 2 (PS1 DR2) and the Local Group Galaxies Survey (LGGS) within a radius of $1''$ for further analysis. Then, similar to the method in Section \ref{sec:By the Near-Infrared Color-Color Diagram}, we select the ``good measurement" sources with $r$, $z$, $H$ band photometric errors less than 0.05 to plot $r-z/z-H$ diagram, and we determine the points of the maximum surface density on $r-z/z-H$ diagram in steps of 0.01 and the range of $r-z$ from 0 to 2.5. Then the piecewise function is used to fit those points. When $r-z<=1.1$, the sigmoid function, which is also known as the Logistic function, is used for fitting, otherwise, the function form is a quadratic curve. As shown in Figure \ref{fig:PS1-Criteria}, we move up the piecewise functions by 0.12 and 0.10 mag for M31 and M33 and take them as the dividing lines between giants and dwarfs. The summary of PS1 criteria are listed in Table \ref{tab:dividing lines}.

 \item \textbf{The $B-V/V-R$ diagram.}

\citet{1998ApJ...501..153M} proved that the $B-V/V-R$ diagram can separate RSGs from the foreground dwarfs. The low surface gravity RSGs is redder in $B-V$ at given $V-R$ than the high surface gravity dwarfs because the $B$ band covers many metallic lines to become sensitive to surface gravity. Although \citet{2009ApJ...703..420M,2012ApJ...750...97D,2016ApJ...826..224M} already determined the dividing line to separate RSGs in M31 and M33 from the foreground dwarfs, we still renew it to derive a quantitative relation of $B-V$ and $V-R$ for dwarfs stars like in previous calculations for $J-H/H-K$ and $r-z/z-H$. The method is the same as in Section \ref{sec:By the Near-Infrared Color-Color Diagram}. The stars whose $B,V,R$-band photometric errors less than 0.05\,mag are chosen to define the borderline by the points of the maximum surface density on the $B-V/V-R$ diagram in step of 0.01 in the range of $V-R$ from -0.25 to 1.30. The Logistic function is taken to fit these points and shifted up by 0.12 and 0.10 mag for M31 and M33 respectively, respectively, as the dividing lines shown in Figure \ref{fig:LGGS-Criteria}. The function forms and coefficients of dividing lines are listed in the third rows of Table \ref{tab:dividing lines}.

The LGGS criteria work very well for bright sources when \citet{2009ApJ...703..420M,2016ApJ...826..224M} limit the sources by $V$ brighter than 20 mag, i.e. the photometric error less than 0.01. But for faint sources with slightly large photometric error, to distinguish dwarfs from giants is very hard. As shown in the left panel of Figure \ref{fig:check_bvr_foreground}, a large portion of the foreground dwarfs selected by the NIR CCD locate in the $B-V/V-R$ region of dwarfs, confirming the consistency between optical and NIR criteria. But when the photometric error increases to 0.05 mag, many of the NIR-selected foreground dwarfs fall into the giants region of the LGGS diagram as shown in the right panel of Figure \ref{fig:check_bvr_foreground}, needless to say that many of the objects have photometric uncertainty of $\sim 0.1\,\mathrm{mag}$. This can be understood that the difference of giants and dwarfs in the $B-V/V-R$ diagram is too small to tolerate any significant photometric error. Comparing Figure \ref{fig:LGGS-Criteria} with Figure \ref{fig:UKIRT-Criteria} and Figure \ref{fig:PS1-Criteria} demonstrates clearly that the difference of colors in optical (mostly $\sim 0.1$ mag) is much less significant than in near-infrared (mostly $\sim 0.2$ mag).  The cross-match of the LGGS catalog with UKIRT by a radius of 1\arcsec results in 92,441 (7.4\% of the initial NIR sample) and 45,279 (22.3\%) associations. These associations are  a small part of the sample, and taking the risk of mixing dwarfs and giants together into account, the LGGS criteria are not used to remove dwarfs in this work.

\end{itemize}

%So when the photometric error is large, there is a risk that dwarfs and giants stars will be confused on $B-V/V-R$ diagram. In addition, we note that not all the sources that have been only removed by Gaia criteria (foreground giants) locate in the giants region of the LGGS criteria. It implies that the LGGS criteria have the possibility to identify giants incorrectly.

\subsection{Removing the Foreground Giants} \label{sec:Removing the Foreground Giants}

It is worth noting that our UKIRT and PS1 criteria can be used only to remove foreground dwarfs, but foreground giant stars cannot be removed in this way. Instead we make use of the Gaia astrometric information to remove them because the foreground giants should present measurable motions.

Together with some foreground dwarfs, the foreground giants are searched by parallaxes and proper motions from Gaia/DR2. Stars are considered to be foreground objects if they satisfy either parallax or proper motion constraints. Specifically, we remove the sources whose distances are less than the Milky Way scale (i.e. 50 kpc; \citealt{2017RAA....17...96L}) with astrometric solution relative error less than 20\% (i.e. $|\sigma_{\omega}/\omega|$, $|\sigma_{\mu_{\alpha*}}/\mu_{\alpha*}|$ and $|\sigma_{\mu_{\delta}}/\mu_{\delta}|$ are both smaller than 20\%). The distances here are calculated with Smith-Eichhorn correction method from Gaia-measured parallax and its error \citep{1996MNRAS.281..211S}. Besides, a source is also removed if the measured proper motion is greater than that expected for a star with a velocity of 500km/s at the distance of M31 and M33, i.e. 0.2mas/yr, and the astrometric solution relative error smaller than 20\%:
\begin{equation} \label{eq:pm_ra}
	\left| \mu_{\alpha*} \right| > 0.2 \mathrm{mas\ yr^{-1}} + 2.0\sigma_{\mu_{\alpha*}},
\end{equation}
\begin{equation} \label{eq:pm_dec}
	\left| \mu_{\delta} \right| > 0.2 \mathrm{mas\ yr^{-1}} + 2.0\sigma_{\mu_{\delta}},
\end{equation}
where $\mu_{\alpha*}$ and $\mu_{\delta}$ are proper motions in right ascension and declination.

The sources only removed by Gaia criteria are expected to be foreground giants. Among the Gaia-UKIRT cross-identified 83,260 (M31) and 18,901 (M33) stars, 14,795 and 2,433 stars are removed by the Gaia criteria as well as by UKIRT. There are 15 and 5 sources identified as foreground stars only by Gaia. However, the stars removed only by Gaia are faint in the $K$ band, mostly fainter than 14 mag. A typical red giant star is as bright as -4 mag in $K$, and 14 mag means a distance of 40 kpc. Thus these sources are very unlikely to be foreground red giants. We tend to believe that the astrometric information of Gaia for most of these sources are uncertain. As these stars actually have the color-magnitude similar to RSGs in the host galaxies, this uncertainty may be caused by the variation of photo-center due to large-scale convection of RSGs \citep{2011AandA...528A.120C}. Therefore, the Gaia-only sources are not removed from the sample, which means no foreground red giants are removed from the sample. This result confirms the argument that the contamination of foreground red giants is very small towards the sightlines of M31 and M33 by \citet{2016ApJ...826..224M}.

%As shown in Figure \ref{fig:check_jhk_foreground} and \ref{fig:check_rzh_foreground}, most of the foreground stars fall in the dwarfs region of UKIRT criteria and PS1 criteria. Some of the removed stars fall in the giants region of UKIRT criteria and PS1 criteria. As we expected, most of these sources are foreground giants only removed by Gaia criteria, and the other sources locate at the bifurcation point of dwarfs and giants as shown in Figure \ref{fig:check_rzh_foreground}, so it may be difficult to separate them well.

\subsection{Comparison with the Besan\c{c}on Model} \label{sec:Comparison with the Besancon Model}

\begin{figure}
	\centering
    \includegraphics[scale=0.45]{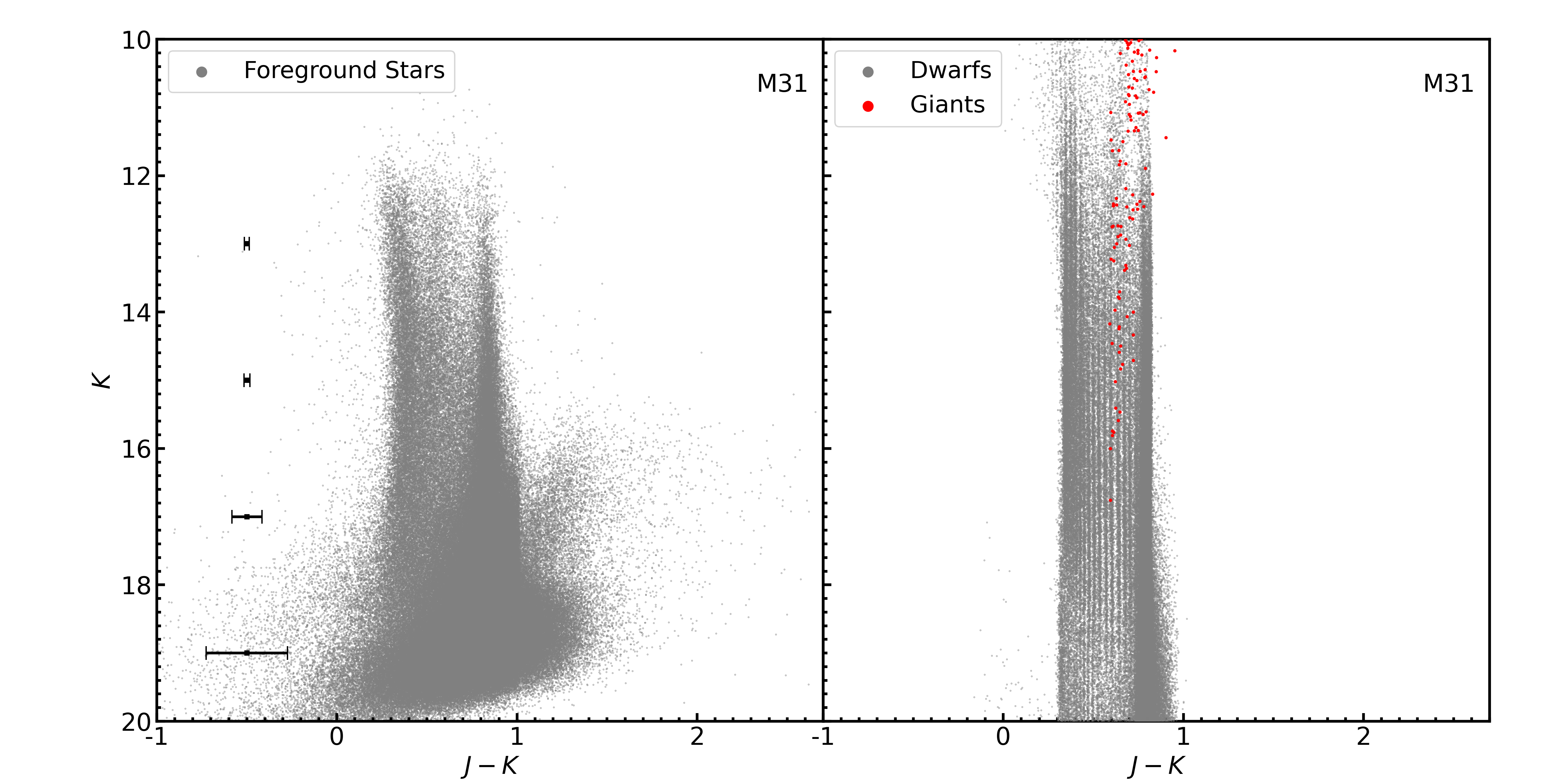}
	\caption{The color-magnitude diagram of the removed foreground stars (left) is compared with the Besancon model (right) for M31. The error bars show the mean error of $J-K$ at different $K$ magnitudes. \label{fig:Besancon_Foreground_M31}}
\end{figure}

\begin{figure}
	\centering
    \includegraphics[scale=0.45]{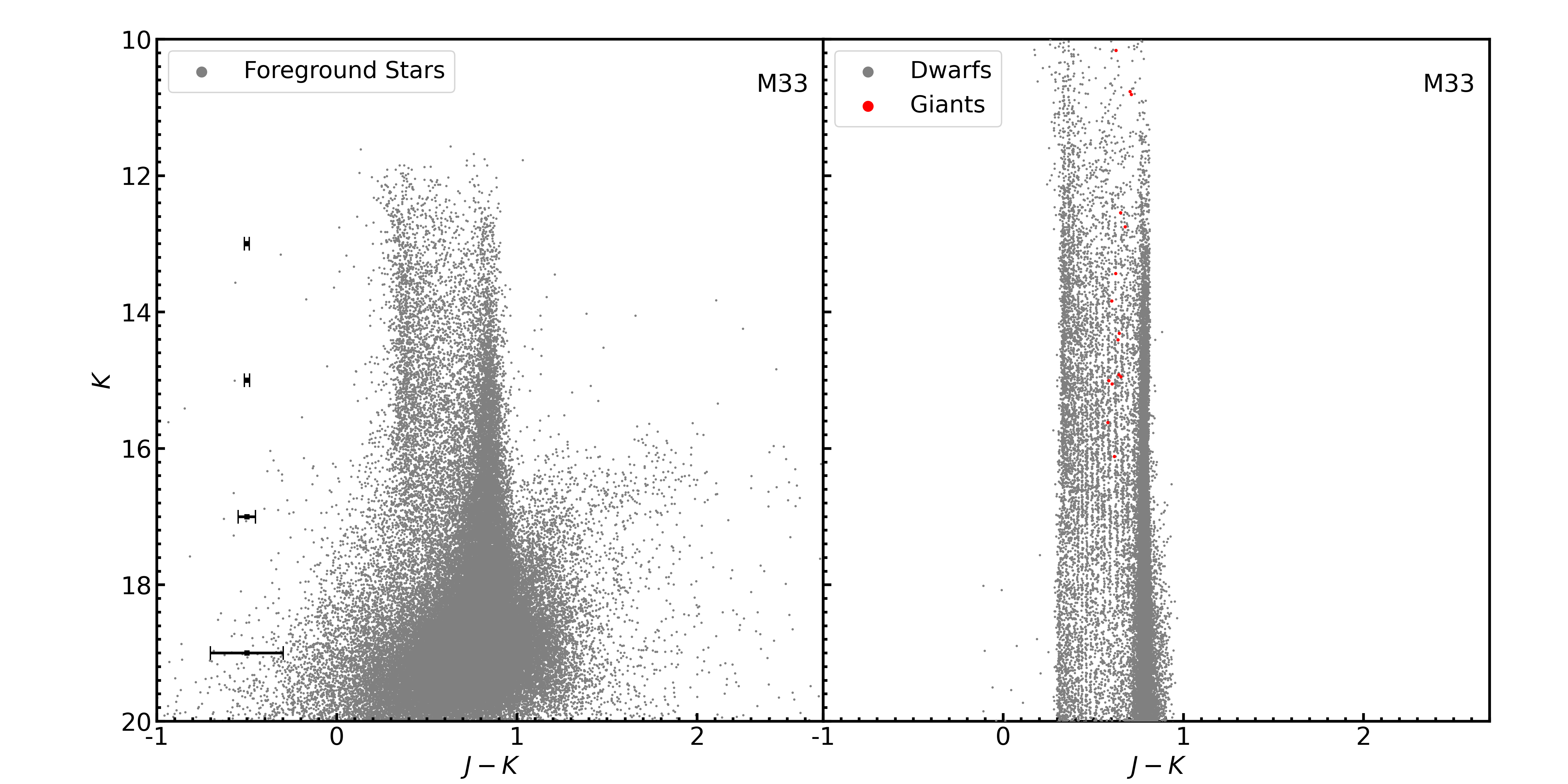}
	\caption{Same as Figure \ref{fig:Besancon_Foreground_M31}, but for M33. \label{fig:Besancon_Foreground_M33}}
\end{figure}

The Besan\c{c}on Milky Way stellar population synthesis model \citep{2003A&A...409..523R} is introduced to examine whether the foreground stars are removed correctly. Within the distance range of 0-50 kpc, the expected foreground stars at the direction of M31 and M33 with a sky area of 12 $\mathrm{deg}^{2}$ and 3 $\mathrm{deg}^{2}$, approximate to the coverage of the UKIRT data, are computed by the Besan\c{c}on model. Their distribution in the $J-K/K$ diagram is displayed in Figure \ref{fig:Besancon_Foreground_M31} and \ref{fig:Besancon_Foreground_M33}, where the red giants (red dots) are diagnosed by $\mathrm{log}\ g < 2.5$ (from G5 later on) since the blue (super)giants with large $\mathrm{log}\,g$ are few and might be removed by our criteria. For the M31 area, the number of foreground giants and foreground dwarfs with $K$ band magnitude brighter than 20 mag and fainter than 12 mag is 55 and 214,157 respectively. For the M33 area, the numbers are 12 and 26,046, respectively. In comparison, our criteria removed 422,741 and 79,176 dwarfs in the sky area of M31 and M33, which is basically consistent with but still more than that predicted by the Besan\c{c}on model. On the contrary, we removed no foreground red giants, which can be understood that the foreground red giants are so bright that they should be saturated in the UKIRT observation with stellar classification flag of -9 (possibly saturated objects) and have been excluded from the initial sample.

%From the results of the Besan\c{c}on model, it can be seen that the proportion of the giants in the foreground stars is very small ($<1\%$). In addition, as shown in Figure 1, those faint RSGs will be seriously contaminated by the foreground stars. Therefore, in order to obtain a pure sample of RSGs, the contamination of foreground stars, especially the foreground dwarfs, must be removed.
%

%%in which only the source removed by Gaia criteria is considered as the foreground giants, which is 1,384 and 908 respectively. This number is consistent with the order of magnitude predicted by the Besan\c{c}on model.

As shown in Figure \ref{fig:Besancon_Foreground_M31} and \ref{fig:Besancon_Foreground_M33}, the foreground dwarfs removed by the UKIRT and PS1 criteria are consistent with the Besan\c{c}on model, but those faint sources in the $K$ band have larger dispersions than the ``no-error" Besan\c{c}on model due to the affect of photometric error. Meanwhile, some stars removed appear on the red side with $J-K >1.0$, which is not present in the Besan\c{c}on model. Re-examination found that these sources are removed by the $r-z/z-H$ criterion, and they are mostly located close to the center of the galaxy whose photometry suffers relatively large uncertainty. Nevertheless, they have little effect on the RSGs sample because they are too red at given brightness for RSGs, and they may be AGB or RGB stars.

The number of foreground stars removed by the three methods is summarized in Table \ref{tab:foreground stars removed by three methods}. Apparently the NIR color-color diagram works much more effectively than the other two methods because the distinction of giants from dwarfs is significant and RSGs are bright in near-infrared.

\section{Identifying RSGs in the $J-K/K$ Diagram} \label{sec:Identifying the RSGs in the CMD}

\subsection{Various Types of Evolved Stars in the $J-K/K$ Diagram} \label{sec:Various Types of Evolved Stars in the CMD}

\begin{figure}
	\centering
    \includegraphics[scale=0.45]{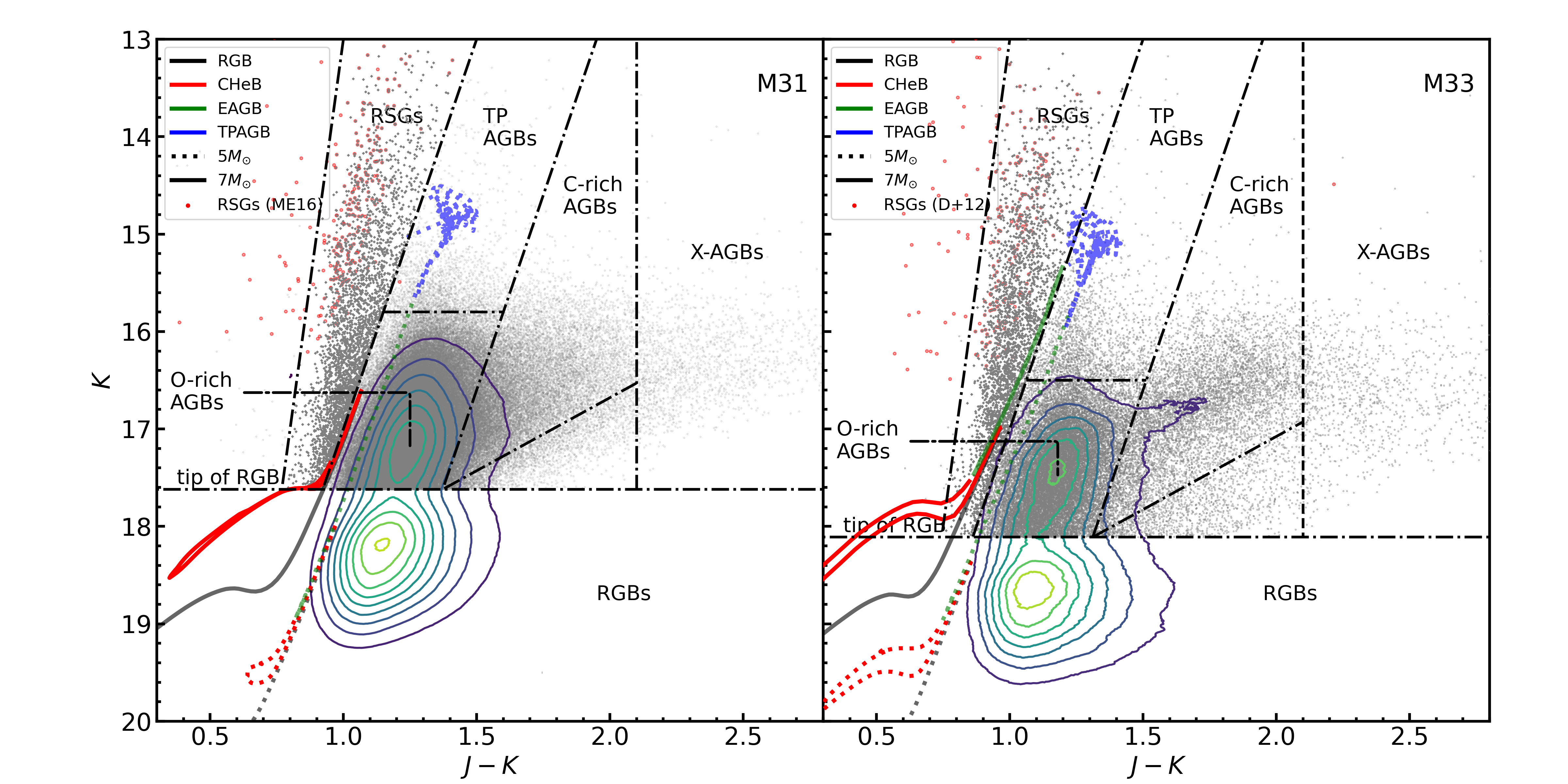}
	\caption{The color-magnitude diagram of the good-measurement member stars with ``N\_Flag=3" and  $\sigma_{JHK} < 0.1$. The divisions of RSGs, TP-AGBs, O-AGBs, C-AGBs, and X-AGBs are indicated by dash lines. The MIST model tracks for a $5M_{\odot}$ and a $7M_{\odot}$ star are illustrated with evolutionary stages including core-helium burning, early-AGB and TP-AGB.   \label{fig:stellar_population}}
\end{figure}

Although this work focuses on the catalogs of RSGs in M31 and M33, other evolved populations can be identified using the CMDs of the member stars after removing the foreground stars, which helps to identify RSGs.  For this purpose, we choose the sources with ``N\_Flag=3" and $JHK$ photometric errors less than 0.1\,mag to define the regions of various populations in the near-infrared CMDs shown in Figure \ref{fig:stellar_population}, and then apply the criteria to classify other less-accurately-measured sources. In order to ensure that the objects are point sources, we drop  those with ``N\_Flag=2" (i.e. only two of $JHK$ bands are marked as point source) but marked as extended source by PS1. On this basis, we define the sources with ``N\_Flag=2'' as ``Rank 2'' sources and sources with ``N\_Flag=3'' as ``Rank 1'' sources.

The division of various populations in the CMD is mainly guided by previous studies of evolved populations in SMC and LMC by \citet{2019AandA...629A..91Y, 2020AandA...639A.116Y, 2020.LMC.Y}. Additionally, the MIST (MESA Isochrones and Stellar Tracks) model \citep{2011ApJS..192....3P,2013ApJS..208....4P,2015ApJS..220...15P,2016ApJS..222....8D,2016ApJ...823..102C} of a 5$M_{\odot}$ and 7$M_{\odot}$ star is referenced for M31 and M33 with  $[\mathrm{Fe/H}]=0.3$ and $[\mathrm{Fe/H}]=0.1$ respectively. Above these empirical and theoretical results, the division is decisively determined by the density contour of the stars in the $J-K/K$ diagram. First, the position of Tip-RGB (TRGB) is determined, which will be discussed in Section \ref{sec:Distance to M31 and M33}, and the sources fainter than TRGB are RGB stars. The sources brighter than TRGB are divided into two major parts: relatively bluer RSGs and relatively redder AGB stars. The former is the leading role to be discussed in details in the following. AGB stars are subdivided into oxygen-rich AGBs (O-rich AGBs), carbon-rich AGBs (C-rich AGBs), extreme AGBs (X-AGBs), and thermally pulsing AGBs (TP-AGBs). The locations of O-rich, C-rich and X-AGB stars in the $J-K/K$ diagram have been previously identified by several works, such as \citet{2008A&A...487..131C} and \citet{2011ApJ...727...53Y}, and our results agree with theirs. But the location of TP-AGB stars is suggested for the first time. It can be seen from Figure \ref{fig:stellar_population} that the TP-AGB branch is almost parallel to and very close to but slightly redder than the RSG branch, which led to previously classifying them into RSGs. However, there is a clear gap from RSGs, and they coincide very well with the 5$M_{\odot}$ MIST TP-AGB model. We expect spectroscopy and light-variation of these objects (c.f. \citealt{2020ApJ...898...24R}) would further confirm the nature of these sources.

The total numbers of different stellar populations are listed in Table \ref{tab:numbers}. Also listed are the numbers of ``Rank 1'' sources in brackets, whose type flags are all ''stellar'' in the $JHK$ bands, implying the lower limit of the number of the sources of each type. As will be shown in Section \ref{sec:Distance to M31 and M33}, the TRGB of M31 and M33 is 17.62 mag and 18.11 mag in $K$ band, which is brighter than the completeness magnitude (17.94 and 18.22). The samples of AGB and RSG stars should be more or less complete, while the samples of RGB stars are incomplete for the fainter ones are not all detectable. M31 is similar to our Milky Way galaxy in type and metallicity and even size, these numbers should be valuable reference for studying the evolved stellar populations in our Galaxy as a whole.

\subsection{RSGs} \label{sec:RSGs}
The red supergiant branch is very obvious in the $J-K/K$ diagram after removing foreground stars shown in Figure \ref{fig:fgd_member.png}. The magnitude and color criteria can then be used to define the RSGs region in this diagram. As pointed out in previous section, RSGs stand out in the $J-K/K$ diagram above the TRGB. We take the $K$ band magnitude of TRGB (17.62 mag for M31 and 18.11 mag for M33) as the lower limit of RSGs. Although this limit may be questioned, we have two reasons. One is that the configuration in the $J-K/K$ diagram is continuous for the RSG branch until the TRGB where the break occurs at the lowest density. The other is that the core helium burning stage of the 7$M_{\odot}$ star in the MIST track starts from about this position, which indicates the lower mass limit of RSG to be about 7$M_{\odot}$.

Here we take two straight lines as the red and blue boundaries to enclose the RSG branch. The quantitative forms of the blue and red boundaries are:
\begin{equation} \label{eq:blue boundaries}
	\mathrm{blue\ boundary}: K=-20.00(J-K)+33.00,\\
\end{equation}
\begin{equation} \label{eq:red boundaries}
	\mathrm{red\ boundary}: K=-8.00(J-K)+25.00.
\end{equation}

%\textbf{Though some high-luminosity RSGs may evolve back to higher temperatures to the left of the blue boundary due to significant mass loss \citep{2016ApJ...825...50G}, the fraction of such sources should be very small, and will not affect the completeness of the RSGs.}

Previous samples of RSGs in M31 and M33 were obtained from optical observation based on the $B-V/V-R$ diagram \citep{2009ApJ...703..420M,2012ApJ...750...97D}. These photometrically classified RSGs were further checked by radial-velocity information for membership determination.  \citet[ME16 for short]{2016ApJ...826..224M} and \citet[D+12 for short]{2012ApJ...750...97D} identified 255 RSGs in M31 and 204 RSGs in M33 with 189 rank1 highly likely supergiants and 15 rank2 possible supergiants. Among them, 240 in M31 and 201 in M33 radial-velocity confirmed RSGs are in our initial sample, of which 180 and 154 sources are considered as member stars and finally 180 and 147 sources are also identified as RSGs in our work. These stars are labelled as ``D" in the column ``LGGSType" of Table \ref{tab:RSGs_M31} and \ref{tab:RSGs_M33}.

These previously identified RSGs in M31 (240) and M33 (201) are compared with the whole sample in Figure \ref{fig:stellar_population}. It can be seen that previous samples miss the red and faint RSGs and include some objects too blue or too red to be a RSG.

\subsection{Sample of RSGs} \label{sec:Number of RSGs}
\subsubsection{Completeness and pureness} \label{sec:Completeness and pureness}

As mentioned in Section \ref{sec:data}, the sources brighter than 17.94 and 18.22 in the $K$ band are complete. With the lower limits of RSGs in M31 and M33 being $K=$17.62 and 18.11, the samples are considered to be complete, i.e. there are 5,498 and 3,055 RSGs in M31 and M33 respectively. RSGs in M31 and M33 are listed in Table \ref{tab:RSGs_M31} and \ref{tab:RSGs_M33} with RA and Dec coordinates, magnitudes, magnitude errors, astrometric information, etc. One thing to be noted is that our preliminary selection of stars is not very strict in that the stars labelled ``-7" for bad pixel are included and that only two of the $JHK$ bands labelled ``stellar" are required. If all the $JHK$ three bands are required to be ``stellar", then the number of RSGs is reduced to 3,268 and 2,804 for M31 and M33 respectively. The number of reduction for M31 is significant and very minor for M33, again due to the much more crowded field in M31. We think these numerals must have underestimate the sample. Moreover, if the sources with two of the three bands labelled ``-7" are removed, the number of RSGs is further reduced to 3,154 and 2,635, respectively.

Comparing the location of various types of evolved stars in the $J-K/K$ diagram with the limiting magnitude of 2MASS, it can be inferred that some RSGs and AGBs are detectable by 2MASS, but not all of them. If the quality flag is further restricted to ``AAA'', then only bright AGBs and RSGs in M31 with $K<16$ can be recognized by the 2MASS photometry. For M33, the situation is reduced because the main 2MASS point source catalog has a brighter limiting magnitude of $K\sim15$ for the ``AAA'' sources.

%For M31, 2MASS data are obtained from 2MASS 6x point source catalog with limit magnitude of $K\sim18.5$. If we further require photometric quality flag to be ``AAA", only bright AGBs and RSGs with $K<16$ can be observed by 2MASS survey. In the case of M33, 2MASS data are obtained from main 2MASS point source catalog with limit magnitude of $K\sim16.5$, with restrict photometric quality flag, only RSGs with brightness greater than $K\sim15$ can be detected by 2MASS survey.

\begin{figure}
	\centering
    \includegraphics[scale=0.4]{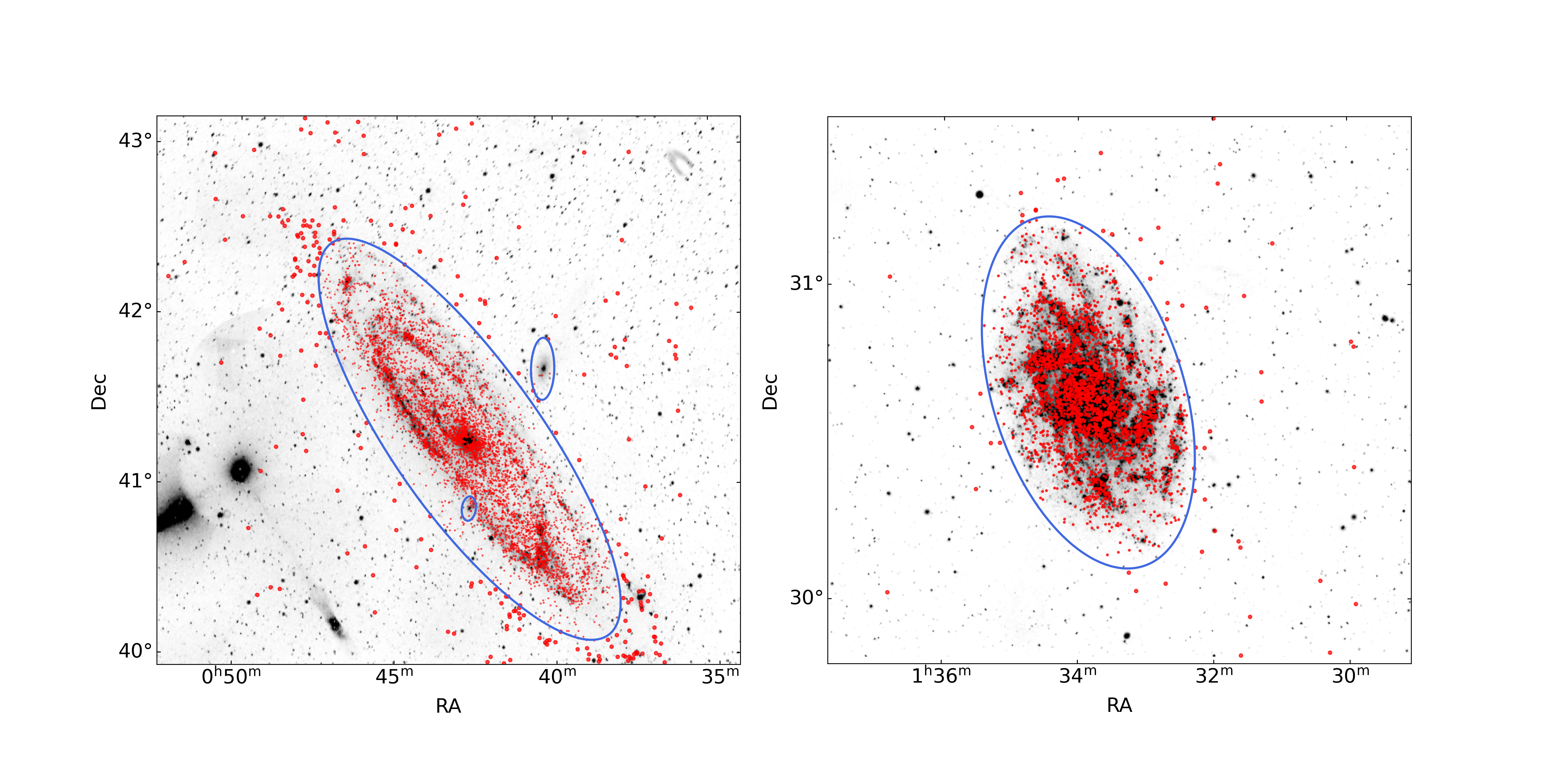}
	\caption{Spatial distribution of RSGs in M31 (left) and M33 (right). The background image is taken from the GALEX ultraviolet observation.  \label{fig:distribution_rsgs}}
\end{figure}

The pollution rate of the RSG sample is estimated from the control field. Firstly the UKIRT fields are divided into geometric regions of the M31, M32, M110 and M33 and the control fields. The region of a galaxy is demarcated by an ellipse whose major and minor axis are determined by the $B=25\ \mathrm{mag/arcsec}^{2}$ isophotes in the $B$ band shown in Figure \ref{fig:distribution_rsgs}. The fields beyond the geometric regions of the galaxies are classified into control fields.  The major axes, minor axes and position angles of the galaxies are listed in Table \ref{tab:diameters}.

The sky area of galaxies is calculated by $S = \pi ab$, where $a$ and $b$ are the half major and half minor axis. The sky area of the control fields is the total sky area minus the sky area of the galaxies. The sky area of different regions and the number of RSGs are listed in Table \ref{tab:diameters}. If all the RSGs in the control fields are foreground stars, or fake RSGs, the pollution rate can be estimated. First, the fake RSGs per sky area is calculated by
\begin{equation} \label{eq:fake rsgs per sky area}
	\Sigma_\mathrm{fake} = N_\mathrm{control\ fields}^\mathrm{RSGs} / S_\mathrm{control\ fields},
\end{equation}
where $N_\mathrm{control\ fields}^\mathrm{RSGs}$ is the number of RSGs in the control fields and $S_\mathrm{control\ fields}$ is the sky area of the control fields.
Then the number of fake RSGs in the region of a galaxy is
\begin{equation} \label{eq:number of fake rsgs}
	N_\mathrm{fake} = \Sigma_\mathrm{fake} \times S_\mathrm{galaxy},
\end{equation}
where $S_\mathrm{galaxy}$ is the sky area of the galaxy and $N_\mathrm{fake}$ is the number of fake RSGs. For M31, $S_\mathrm{galaxy} = S_\mathrm{M31} + S_\mathrm{M110}$ since M32 is inside M31. The pollution rate is then
\begin{equation} \label{eq:pollution rate}
	P = N_\mathrm{fake} / N_\mathrm{galaxy},
\end{equation}
where $N_\mathrm{galaxy}$ is the number of RSGs within the galaxy's area. For M31, $N_\mathrm{galaxy} = N_\mathrm{M31} + N_\mathrm{M32} + N_\mathrm{M110}$. The pollution rate turns out to be 1.30\% and 0.49\% for the catalog of RSGs in  M31 and M33 respectively.  Indeed the fake RSGs in the control fields locate around the rim of the galaxies as shown in Figure \ref{fig:distribution_rsgs}, which indicates some of them are actually the members of M31 or M33. In another word, M31 and M33 extend to a larger area than the labelled ellipse defined by $25\ \mathrm{mag/arcsec}^{2}$ isophotes. The true pollution rate should be smaller than the above derived values.

\subsubsection{Spatial Distribution} \label{sec:Spatial Distribution}

The spatial distribution of the selected RSGs is shown in Figure \ref{fig:distribution_rsgs} with the GALEX ultraviolet image as the background to display the massive star regions. The locations coincide very well with the spiral arms, which are expected for RSGs as massive stars. This structure supports our identification of RSGs.

\subsubsection{Density of RSGs as a Function of Metallicity} \label{sec:Number of RSGs as a Function of Metallicity}

\begin{figure}
	\centering
    \includegraphics[scale=0.7]{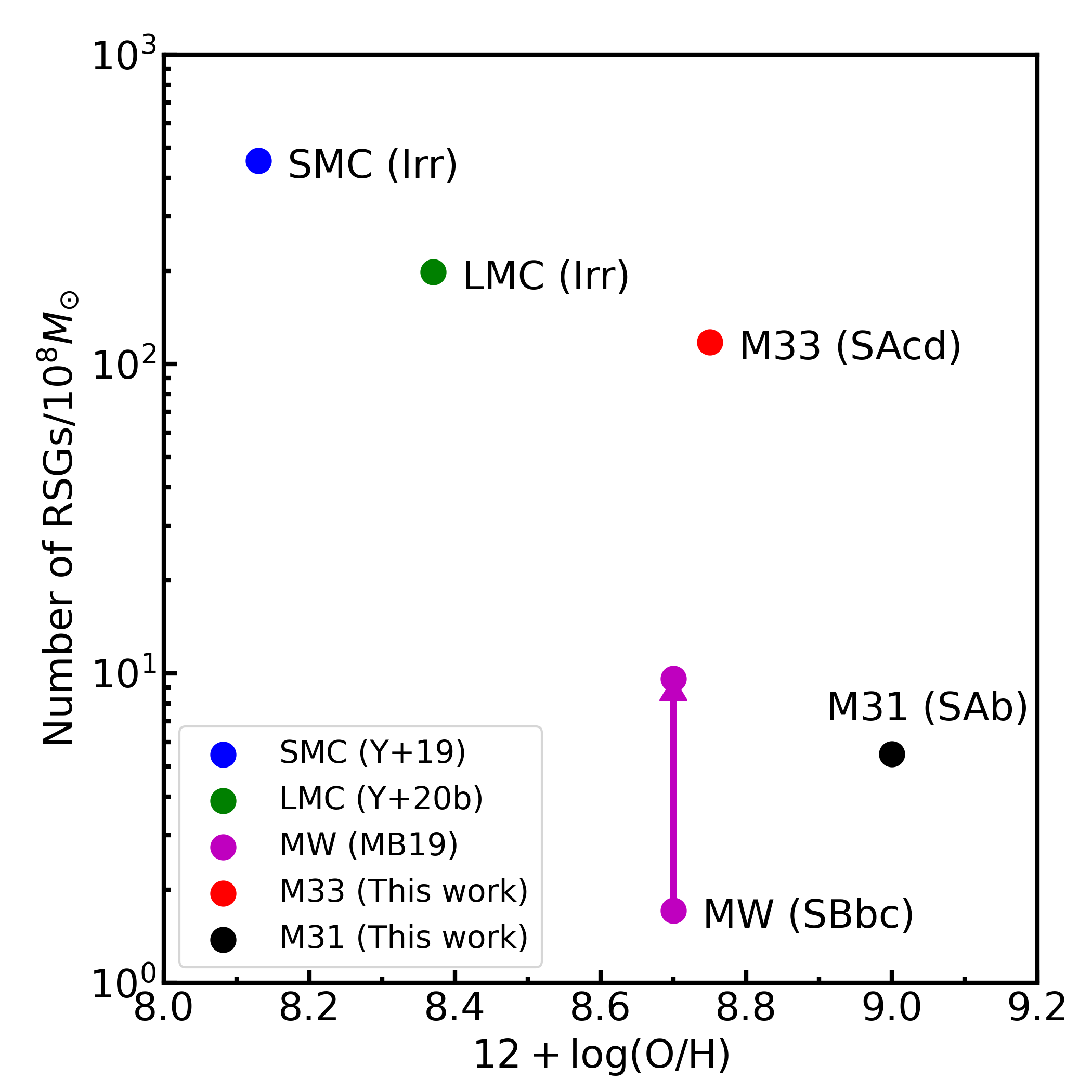}
	\caption{Variation of the number of RSGs per stellar mass with metallicity for five galaxies. For the Milky Way, the lower limit comes from the already identified number of RSGs by \citet{2019AJ....158...20M} while the higher value from the predicted number by \citet{1989IAUS..135..445G}. \label{fig:number_metallicity}}
\end{figure}

The number of RSGs is 5,498 within 2.566 deg$^2$ of M31 and 3,055 within 0.644 deg$^2$ of M33.  Apparently the surface density of RSGs is not the same. It is well known that metallicity influences the time of a massive star to spend in the RSG stage.  In order to characterize the density and the massive star formation rate, the number of RSGs is normalized to stellar mass of the galaxy. In addition to M31 and M33, we supplement the other galaxies (SMC, LMC and MW) whose RSG samples are systematically studied. The number of RSGs in SMC and LMC is 1,405 and 2,974 from \citet[Y+19 for short]{2019AandA...629A..91Y} and \citet[Y+20b for short]{2020.LMC.Y}. It should be noted that the sample of RSGs in the Milky Way galaxy is incomplete because of the interstellar extinction itself and the complexity caused by it. The recent search in the Galaxy found 889 nearby RSGs \citep[MB19 for short]{2019AJ....158...20M}, meanwhile \citet{1989IAUS..135..445G} predicted at least 5,000 RSGs. Both values are displayed in Figure \ref{fig:number_metallicity}, and the former one indicates the lower limit. The adopted stellar masses are $3.1\times10^{8} \rm{M_{\odot}}$, $1.5\times10^{9}\rm{M_{\odot}}$, $5.2\times10^{10}\rm{M_{\odot}}$, $2.6\times10^{9}\rm{M_{\odot}}$ and $1.0\times10^{11}M_{\odot}$ for SMC, LMC, MW, M31 and M33 respectively \citep{2015arXiv151103346B,2016MNRAS.457..844F}.  The metallicity from SMC and LMC to M33 and M31 is increasing, from sub\-solar metallicity to super\-solar metallicity. Specifically,  the value of $12+\log(\mathrm{O/H})$ is 8.13 \citep{1990ApJS...74...93R}, 8.37 \citep{1990ApJS...74...93R}, 8.70 \citep{1995RMxAC...3..133E}, 8.75 \citep{1997ApJ...489...63G}, 9.00 \citep{1994ApJ...420...87Z} for SMC, LMC, MW, M33 and M31.

The RSG density per $10^8 M_{\odot}$ is presented in Figure \ref{fig:number_metallicity}.   With increasing metallicity, the RSG density per stellar mass decreases rapidly.  When the metallicity $12+\log(\mathrm{O/H})$ increases by 0.9 dex, the number of RSGs per stellar mass decreases by about 60 times. This can be understood that metallicity affects the time of stars spent in different evolutionary stages. Previous studies have shown that metallicity affects the ratio of blue-to-red supergiants (B/R) and Wolf-Rayet stars to RSGs (WR/RSGs). When metallicity increases by 0.9 dex, the B/R ratio and the W-R/RSG ratio increase by about 7 times \citep{1980A&A....90L..17M} and 100 times \citep{2002ApJS..141...81M} respectively. Our conclusion that the density of RSGs decreases with increasing metallicity is consistent with this scenario. On the other hand, metallicity cannot be the only factor to influence the density of RSGs. M33 is much more metal-rich than LMC, but holds very similar RSG density. In fact, M33 is the only SAcd-type galaxy in them, the strong star-forming activity may account for higher massive star formation rate. Among these galaxies, M31 should be the most similar one to our galaxy in both type and metallicity. Taking M31 as the reference, the number of RSGs in our galaxy should be $\sim 2,800$ after scaling by stellar mass. But this is only the lower limit of the number of RSGs in our galaxy. Because the metallicity of the Milky Way is lower than that of M31, the number of RSGs per stellar mass should be higher than that of M31. As shown in Figure \ref{fig:number_metallicity}, the \citet{1989IAUS..135..445G} prediction of the number of 5,000 RSGs in our galaxy agrees well with the overall trend of the number of RSGs per stellar mass with metallicity, but is still lower than the overall trend. A detailed study of this problem will be presented in our future work.

\subsection{Tip-RGB of M31 and M33} \label{sec:Distance to M31 and M33}
\begin{figure}
	\gridline{\fig{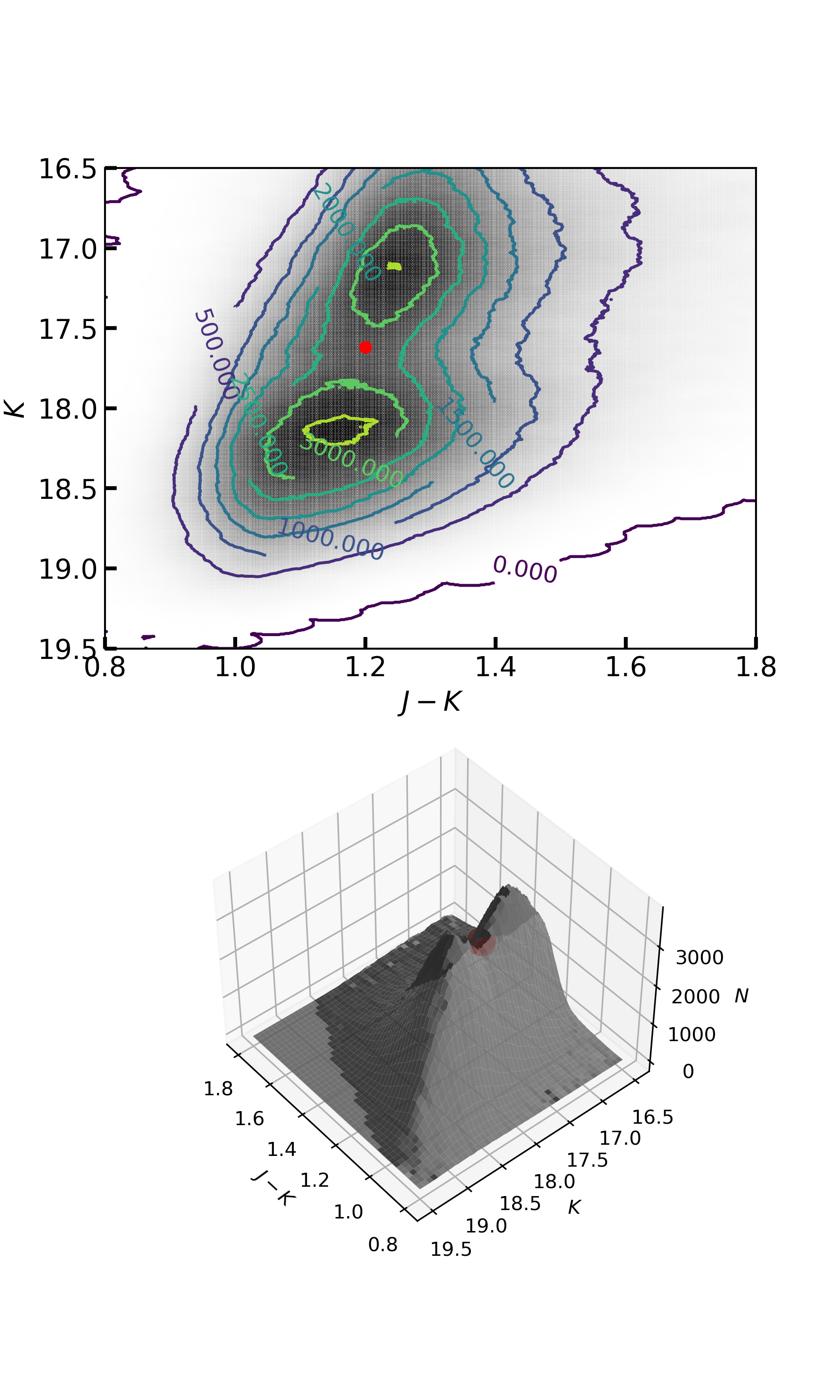}{0.55\textwidth}{(a)}
		  	\fig{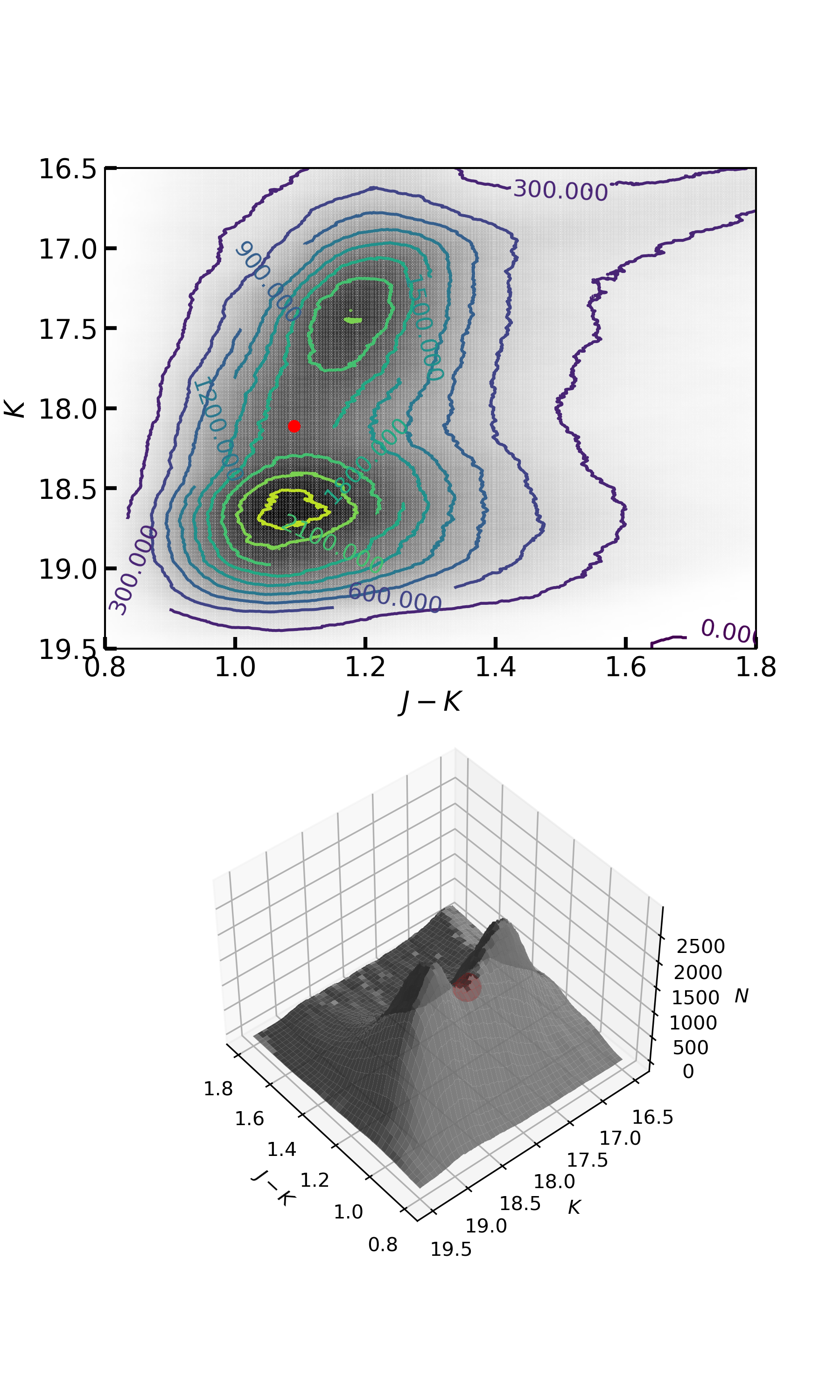}{0.55\textwidth}{(b)}}
	\caption{Saddle point on the surface density map of $J-K/K$ by the red dot (top). The bottom panel is a three-dimensional diagram, where the height indicates the surface density. \label{fig:trgb}}
\end{figure}

The photometry depth of UKIRT covers the tip of the red giant branch (TRGB). In Figure \ref{fig:stellar_population},  the density of stars decreases gradually and then increases in the $J-K/K$ diagram from RGB to AGB, and the location of TRGB is at the lowest density.  In mathematics, the saddle point needs to satisfy two conditions, one is that this point is a stationary point and can be expressed by
\begin{equation} \label{eq:stationary point}
	\frac{\partial N}{\partial X} = 0, \frac{\partial N}{\partial Y} = 0,
\end{equation}
where $N$ is the surface density, $X$ and $Y$ represent  $J-K$ and $K$ respectively.
The other condition is that the determinant of Hessian matrix is less than 0 and can be expressed by
\begin{equation} \label{eq:Hessian matrix}
	\mathrm{det}\left(\begin{array}{cc}\frac{\partial^{2} N}{\partial X^{2}} & \frac{\partial^{2} N}{\partial X \partial Y} \\ \frac{\partial^{2} N}{\partial Y \partial X} & \frac{\partial^{2} N}{\partial Y^{2}}\end{array}\right)<0,
\end{equation}
The apparent magnitude and color index of saddle points obtained by this algorithm are $J-K = 1.20$ and $K = 17.62$ for M31, $J-K = 1.09$ and $K = 18.11$ for M33 shown in Figure \ref{fig:trgb}.

In principle, the apparent magnitude depends on the distance and the metallicity of the galaxy, and the color index depends only on the metallicity. This provides a route to derive the metallicity and distance of M31 and M33.  With the increase of metallicity,  $J-K$ of TRGB increases and the $K$-band absolute magnitude of TRGB becomes brighter \citep{2004A&A...424..199B,2018AJ....156..278G}. Using the relation between $K$ band absolute magnitude and $(J-K)_{0}$ of TRGB  from \citet{2018AJ....156..278G}, the absolute magnitude can be calculated:
\begin{equation} \label{eq:TRGB}
	M^\mathrm{TRGB}_{K} = -2.78[(J-K)_{0}-1.0]-6.26.
\end{equation}
After subtracting $E(B-V) = 0.07$ \citep{2000glg..book.....V} for the  Galactic foreground extinction, i.e. $A_{J} = 0.24A_{V}$ and $A_{K} = 0.08A_{V}$ \citep{2019ApJ...877..116W},  the absolute magnitude of TRGB are -6.72 mag and -6.41 mag for M31 and M33 respectively. Then the distance modulus $(m-M)_{0}$ of M31 and M33 are 24.32 and 24.50, which are in general agreement with previous results and slightly smaller than 24.40 for M31 by \citet{2009A&A...507.1375P} and 24.66 for M33 by \citet{2007Natur.449..872O}.

\section{Summary} \label{sec:summary and conclusion}

The archival photometric data taken by UKIRT/WFCAM from mid-2005 to 2008 is used to select RSGs in M31 and M33, which is supplemented by the PS1, LGGS and Gaia photometry and astrometric information. The foreground dwarfs are removed mainly by their obvious branch in the $J-H/H-K$ diagram due to the significant darkening in $H$ at higher effective temperature of dwarfs than giants. This identification of dwarfs in NIR color-color diagram is examined further by optical/infrared colors, specifically in the $r-z/z-H$ and $B-V/V-R$ diagram, and also supported by the Gaia measurement of parallax and proper motion.   The depth of photometry, complete to about $K=18$mag, combined with the criteria limited within the NIR colors, brings about a complete sample of RSGs in M31 and M33. The RSGs are identified in the members' $J-K/K$ diagram from their outstanding locations caused by high luminosity and low effective temperature. The final sample includes 5,498 and 3,055 RSGs in M31 and M33, respectively.  The control fields are used to estimate the pollution rate of the RSGs in our sample, which is found to be less than  1\%. By comparing with LMC, SMC and the MW galaxy, it is found that the number of RSGs per stellar mass decreases with metallicity, which can be understood by the metallicity effect on the duration of the RSG phase for a star. In addition, the type of galaxy may also play a role in that an Sc type hosts more RSGs than an Sb galaxy.

The other evolved low-mass stars are identified  from the $J-K/K$ diagram of the member stars according to empirical divisions and theoretical models. The complete sample is collected for  oxygen-rich AGBs, carbon-rich AGBs, extreme AGBs and thermally pulsing AGBs. The tip-RGB is recognized by the minimum density in the $J-K/K$ diagram, which is $J-K = 1.20$ and $K = 17.62$ for M31, $J-K = 1.09$ and $K = 18.11$ for M33. Its implication on the distance modulus is discussed.

\acknowledgments We are grateful to Profs. Jian Gao, Hai-Bo Yuan and Ms. Yuxi Wang for very helpful discussions and Prof. Roberta M. Humphreys for very good suggestions. We are grateful to Dr. Mike Read for his kind assistance about the UKIRT data. This work is supported by National Key R\&D Program of China No. 2019YFA0405503 and NSFC 11533002. This work has made use of data from UKIRT, PS1, LGGS and Gaia.


\vspace{5mm}
\facilities{}


\software{Astropy \citep{2013A&A...558A..33A},
		  TOPCAT \citep{2005ASPC..347...29T},
		  LMfit \citep{2018zndo...1699739N}
          }




\bibliography{paper}{}
\bibliographystyle{aasjournal}

%% Tables

\begin{longrotatetable}
	\begin{deluxetable*}{ccccc} \label{tab:dividing lines}
		\tablecaption{Function and its coefficients of dividing lines in Figure \ref{fig:UKIRT-Criteria}, \ref{fig:PS1-Criteria} and \ref{fig:LGGS-Criteria}.}
	
		\tablewidth{0pt}
		\tablehead{
			\colhead{} & \colhead{Color criteria} & \colhead{Magnitude criteria} & \colhead{Coefficients} & \colhead{} \\
			\colhead{} & \colhead{} & \colhead{} & \colhead{M31} & \colhead{M33}
		}
		\startdata
		UKIRT & $H-K<0.10$ & All sources & & \\
		{} & $H-K<=0.13$ & $J-H < a(H-K) + b$ & $a=2.265, b=0.329$ & $a=2.837, b=0.258$ \\
		{} & $H-K>0.13$ & $J-H < a(H-K)^{2}+b(H-K)+c$ & $a=-1.384, b=0.241, c=0.616$ & $a=-1.441, b=0.241, c=0.620$ \\
		PS1 & $r-z<0.30$ & All sources & & \\
		{} & $r-z<=1.10$ & $z-H < \frac{KP_{0}e^{m(r-z-X_{0})}}{K+P_{0}(e^{m(r-z-X_{0})}-1)} + C$ & $K=1.265, P_{0}=0.409, m=5.357$ & $K=1.382, P_{0}=0.484, m=5.027$ \\
		{} & {} & {} & $X_{0}=0.106, C=0.712$ & $X_{0}=0.087, C=0.610$ \\
		{} & $r-z>1.10$ & $z-H < a(r-z)^{2}+b(r-z)+c$ & $a=0.143,b=-0.283,c=2.103$ & $a=0.137,b=-0.233,c=2.067$ \\
		LGGS & $V-R<0.50$ & All sources & & \\
		{} & All sources & $B-V < \frac{KP_{0}e^{m(V-R-X_{0})}}{K+P_{0}(e^{m(V-R-X_{0})}-1)} + C$ & $K=2.242, P_{0}=0.434, m=3.547$ & $K=2.572, P_{0}=0.475, m=2.959$ \\
		{} & {} & {} & $X_{0}=0.005, C=-0.461$ & $X_{0}=-0.126, C=-0.666$ \\
		\enddata
	\end{deluxetable*}
\tablecomments{In fact, the coefficients listed in this table are based on the fitting results of the maximum surface density of the dwarf branch, which need to be moved up as the dividing line. The values of shift are described in Section \ref{sec:By the Near-Infrared Color-Color Diagram} and \ref{sec:Double Check by Optical/Infrared Color-Color Diagrams}.}
\end{longrotatetable}

%\begin{deluxetable*}{cccccccc}[ht] \label{tab:ps1 flags}
%	\tablecaption{Rejection Criteria of PS1 DR2 Flag}
%
%	\tablewidth{0pt}
%	\tablehead{
%		\colhead{Column Name} & \colhead{Rejection Criteria of Flag Value}
%	}
%	\startdata
%    XinfoFlag (X is $g$ or $r$) & 8, 16, 256, 1024 \\
%	XinfoFlag2 (X is $g$ or $r$) & 4, 1048576 \\
%	XinfoFlag3 (X is $g$ or $r$) & 1, 2, 4 \\
%	qualityFlag & 1, 2, 64, 128 \\
%	\enddata
%\end{deluxetable*}

\begin{deluxetable*}{ccccccc}[ht] \label{tab:foreground stars removed by three methods}
	\tablecaption{Number of foreground stars removed based on the UKIRT, PS1 and Gaia observations}
	\tablewidth{0pt}
	\tablehead{
		\colhead{} & \colhead{Initial sample} & \colhead{UKIRT} & \colhead{PS1} & \colhead{Gaia} & \colhead{Foreground stars} & \colhead{Member stars}
	}
	\startdata
	M31 & 1,245,930 & 414,490 & 102,174 & 14,836 & 422,741 & 823,189 \\
	M33 & 203,486 & 77,091 & 17,648 & 2,451 & 79,176 & 124,310 \\
	\enddata
\end{deluxetable*}

\begin{deluxetable*}{ccccccc}[ht] \label{tab:numbers}
	\tablecaption{Number of various evolved stellar populations in M31 and M33$^\dag$}
	\tablewidth{0pt}
	\tablehead{
		\colhead{} & \colhead{RGBs} & \colhead{O-rich AGBs} & \colhead{C-rich AGBs} & \colhead{X-AGBs} & \colhead{TP-AGBs} & \colhead{\textbf{RSGs}}
	}
	\startdata
	M31 & 447,942 (325,288) & 257,055 (174,348) & 46,682 (33,708) & 7,037 (5,006) & 3,625 (2,223) & \textbf{5,498 (3,268)} \\
	M33 &  61,862 (48,167) & 36,353 (29,879) & 10,218 (7,972) & 2,713 (1,989) & 2,126 (1,962) & \textbf{3,055 (2,804)} \\
	\enddata
\tablecomments{$^\dag$ The numbers in the brackets include only the sources labelled as ``stellar'' in all the UKIRT/$JHK$ bands. }
\end{deluxetable*}

\begin{deluxetable}{cccccccccc} \label{tab:RSGs_M31}
	\tablecaption{Catalog of red supergiants in M31}
	\tablewidth{0pt}
	\tablehead{\colhead{RA} & \colhead{Dec} & \colhead{JMag} & \colhead{JMag\_Err} & \colhead{HMag} & \colhead{HMag\_Err} & \colhead{KMag} & \colhead{KMag\_Err} & \colhead{...} & \colhead{N\_Flag} \\
	\colhead{(deg)} & \colhead{(deg)} & \colhead{(mag)} & \colhead{(mag)} & \colhead{(mag)} & \colhead{(mag)} & \colhead{(mag)} & \colhead{(mag)} & \colhead{...} & \colhead{}	
	}
	\startdata
	11.446398 & 42.345503 & 16.708 & 0.013 & 15.907 & 0.01 & 15.689 & 0.01 & ... & 3 \\
	11.462842 & 42.367343 & 17.546 & 0.022 & 16.779 & 0.014 & 16.514 & 0.016 & ... & 2 \\
	11.469417 & 42.345959 & 15.273 & 0.01 & 14.484 & 0.01 & 14.077 & 0.01 & ... & 3 \\
	11.518298 & 42.411805 & 16.206 & 0.01 & 15.398 & 0.01 & 15.118 & 0.01 & ... & 3 \\
	11.534664 & 42.378428 & 16.897 & 0.014 & 16.058 & 0.01 & 15.78 & 0.01 & ... & 3
	\enddata
	\tablecomments{(This table is available in its entirety in machine-readable form.)}
\end{deluxetable}

\begin{deluxetable}{cccccccccc} \label{tab:RSGs_M33}
	\tablecaption{Catalog of red supergiants in M33}
	\tablewidth{0pt}
	\tablehead{\colhead{RA} & \colhead{Dec} & \colhead{JMag} & \colhead{JMag\_Err} & \colhead{HMag} & \colhead{HMag\_Err} & \colhead{KMag} & \colhead{KMag\_Err} & \colhead{...} & \colhead{N\_Flag} \\
	\colhead{(deg)} & \colhead{(deg)} & \colhead{(mag)} & \colhead{(mag)} & \colhead{(mag)} & \colhead{(mag)} & \colhead{(mag)} & \colhead{(mag)} & \colhead{...} & \colhead{}	
	}
	\startdata
	23.222635 & 30.17342 & 18.946 & 0.087 & 18.205 & 0.055 & 18.085 & 0.081 & ... & 2 \\
	23.202302 & 30.172458 & 17.763 & 0.033 & 16.989 & 0.018 & 16.809 & 0.026 & ... & 3 \\
	23.042234 & 30.153804 & 18.004 & 0.04 & 17.261 & 0.024 & 17.061 & 0.032 & ... & 3 \\
	23.175996 & 30.051969 & 18.296 & 0.051 & 17.575 & 0.031 & 17.433 & 0.046 & ... & 2 \\
	22.900231 & 29.823271 & 16.11 & 0.01 & 15.437 & 0.01 & 15.15 & 0.01 & ... & 3
	\enddata
	\tablecomments{(This table is available in its entirety in machine-readable form.)}
\end{deluxetable}

\begin{deluxetable*}{cccccccc}[ht] \label{tab:diameters}
	\tablecaption{Regions of M31, M32, M110 and M33 and the number of RSGs in each of them.}
	\tablewidth{0pt}
	\tablehead{
		\colhead{} & \colhead{RA} & \colhead{Dec} & \colhead{$\mathrm{Major\ axis}^\mathrm{a}$} & \colhead{$\mathrm{Minor\ axis}^\mathrm{a}$} & \colhead{$\mathrm{Position\ angle}$} & \colhead{Sky area} & \colhead{Number of RSGs} \\
		\colhead{} & \colhead{(deg)} & \colhead{(deg)} & \colhead{(deg)} & \colhead{(deg)} & \colhead{(deg)} & \colhead{($\mathrm{deg}^{2}$)} & \colhead{}
	}
	\startdata
	M31 & 10.684793 & 41.269065 & 3.176 & 1.029 & $45^\mathrm{b}$ & 2.567 & 5,225 \\
	M32 & 10.674300 & 40.865287 & 0.145 & 0.108 & $-15^\mathrm{b}$ & 0.012 & 25 \\
	M110 & 10.092000 & 41.685306 & 0.365 & 0.183 & $-2.5^\mathrm{b}$ & 0.052 & 6 \\
	M33  & 23.462042 & 30.660222 & 1.180 & 0.695 & $23^\mathrm{a}$ & 0.644 & 3,001 \\
	\enddata
	\tablecomments{
	$^\mathrm{a}$ \citet{1991rc3..book.....D} \\
	$^\mathrm{b}$ \citet{2003AJ....125..525J} \\
	}
\end{deluxetable*}

%% Figures

\clearpage

\end{CJK*}
\end{document}